\documentclass[%
12pt,
reprint,
 superscriptaddress,
 amsmath,amssymb,
 aps,UTF8,
]{revtex4-1}

\usepackage{graphicx}% Include figure files
\usepackage{dcolumn}% Align table columns on decimal point
\usepackage{bm}% bold math
\usepackage{xcolor}
\usepackage{url}
\usepackage{float}
\usepackage{tabularx}
\usepackage{lipsum}
\usepackage{array}

\usepackage{booktabs}
\usepackage{subfigure}
\usepackage{multirow}

\begin{document}

\title{Estimating the production of dark photons with $\eta$ decay in Ultra-peripheral Collisions}%% Force line breaks with \\
%%\thanks{A footnote to the article title}%
\author{Wei Kou}
\email{kouwei@impcas.ac.cn}
\affiliation{Institute of Modern Physics, Chinese Academy of Sciences, Lanzhou 730000, China}
\affiliation{University of Chinese Academy of Sciences, Beijing 100049, China}
\author{Xurong Chen}
\email{xchen@impcas.ac.cn (Corresponding author)}
\affiliation{Institute of Modern Physics, Chinese Academy of Sciences, Lanzhou 730000, China}
\affiliation{University of Chinese Academy of Sciences, Beijing 100049, China}
%\affiliation{Guangdong Provincial Key Laboratory of Nuclear Science, Institute of Quantum Matter, South China Normal University, Guangzhou 510006, China}


\begin{abstract}
	 We propose that the signal of dark photons can be found in the decay channel of $\eta$, and that a certain number of events of dark photon leptonic decay can be observed in the double-photon collision process of heavy-ion ultra-peripheral collisions (UPC). We estimate the total cross-section for producing dark photons in ultra-peripheral $PbPb$ collisions at existing and planned future hadron colliders, as well as the number of signal events. Our results support the search for dark photon signals in ultra-peripheral $PbPb$ collisions. We consider the recent signals of $\eta$ decays produced in LHC $pp$ collisions and estimate the number of dark photon events from the same number of $\eta$ decays from CMS. Based on CMS results, we estimate that there will be at least 1000 signal events of dark photons observed for the mass $m_{A^\prime}$ less than 500 MeV. In this study, we propose a method using $\eta$ meson decay to search for signs of dark photons at heavy-ion colliders, such as LHC and RHIC.
	
\end{abstract}

%% \keywords{Suggested keywords}%Use showkeys class option if keyword
\maketitle

%\tableofcontents

\section{Introduction}
\label{sec:intro}
The high-energy experiments in hadron physics provide an important window for testing the theory of quantum chromodynamics (QCD), the strong interaction theory of the standard model. The study of the properties of the $\eta$ meson provides a test of the basic symmetries of QCD in the standard model, and even the possibility of a theoretical precedent beyond the standard model (BSM). In addition, the study of the $\eta$ meson may provide theoretical basis for anomalous behavior of basic symmetries. The first prediction of the $\eta$ meson came from the Sakata model in the 1950s \cite{sakata1956composite,okun1958some,yamaguchi1958possible,yamaguchi1959model,ikeda1959possibile}, and was finally completed by Gell-Mann and Ne'eman \cite{Gell-Mann:1961omu,ne1961derivation} using the eightfold way to fill in the last piece of the pseudoscalar meson octet. In 1961, the $\eta$ meson was discovered in the three-pion resonance state \cite{Pevsner:1961pa}, but due to isospin symmetry, Gell-Mann did not identify this channel. In addition to the traditional studies of $\eta$ decay modes (see the recent review paper \cite{Gan:2020aco}), $\eta$ also serves as a laboratory for searching for new weakly coupled light particles at the GeV scale. These particles include dark photons and other hidden gauge bosons, light Higgs-like scalars, and axion-like particles. These particle states are predicted to be associated with dark matter and other beyond the BSM frameworks, and have become one of the most active research areas in phenomenology over the past decade \cite{Essig:2013lka,Alekhin:2015byh,Battaglieri:2017aum,Alexander:2016aln}.

Dark matter dominates the cosmic matter density, and its cosmological abundance and stability may indicate the existence of a dark sector with its own forces, symmetries, and spectrum of frequencies, perhaps as rich as the standard model \cite{Beacham:2019nyx}. Meson studies provide a unique opportunity to discover dark matter in the MeV-GeV mass range \cite{Nelson:1989fx,Fayet:2006sp}. From the perspective of symmetry, the symmetries of the standard model are extended at the level of dark matter. In particular, the extension of the SU(3)$\times$SU(2)$\times$U(1) gauge symmetry predicts new light vector bosons - dark photons $A^\prime$. In this way, searching for dark photons in experiments can utilize these couplings to explore the contributions of dark photons in electromagnetic interactions \cite{Boveia:2018yeb,Filippi:2020kii}. Various experiments currently exist to study the properties of dark photons, including references \cite{NA482:2015wmo,KLOE-2:2016ydq,BaBar:2017tiz,LHCb:2017trq,HPS:2018xkw,LHCb:2019vmc,NA62:2019meo,Goncalves:2020czp}. In particular, it has been found that the study of meson decays can provide probes of dark photons, largely due to the significant number of meson events produced by large hadron colliders, which can provide a considerable number of decay modes.

Searching for light hidden particles through the study of $\eta$ decay modes is feasible. In addition to $\eta$ decay being a primary channel for searching for various exclusive hidden particles produced by proton beam dump \cite{Batell:2009di,deNiverville:2011it,Gninenko:2012eq,Alekhin:2015byh,Berlin:2018pwi,Tsai:2019buq}, one of the main reasons is that tagging $\eta$ decay can easily distinguish between different models based on the other particles in final states. Therefore, high-energy hadron colliders, including future ones, are important choices for searching for candidates of dark photons and dark matter \cite{Boveia:2018yeb,Bruce:2018yzs}.

In this work, we propose that in UPC, it may be possible to find the channel where $\eta$ decays into a dark photon and a classical photon, and consider the light decay of the dark photon, ultimately leading to the reconstruction of events with a specific mass of the dark photon. We also discuss that the data from the existing large hadron collider (LHC)'s proton-proton collisions are likely to find significant signals of dark photon decays from $\eta$ production \cite{CMS:2023thf}. The organization of this paper is as follows: In the next section, we briefly introduce some knowledge about the production of dark photons $A^\prime$ from $\eta$ decays, and describe the method of calculating the total cross-section in lead-lead UPC. In the third section, we estimate the dark photon signal events using the recent LHC data \cite{CMS:2023thf}, followed by a discussion of the current status. Finally, we provide a summary and future outlook, including the prospects for the proposed $\eta$ factory and electron-ion colliders in the search for dark photons.

\section{production of dark photons in Ultra-peripheral lead-lead collisions}
\label{sec:form}

\subsection{UPC's cross-section}
\begin{figure}[htbp]
	\centering
	\includegraphics[width=0.45\textwidth]{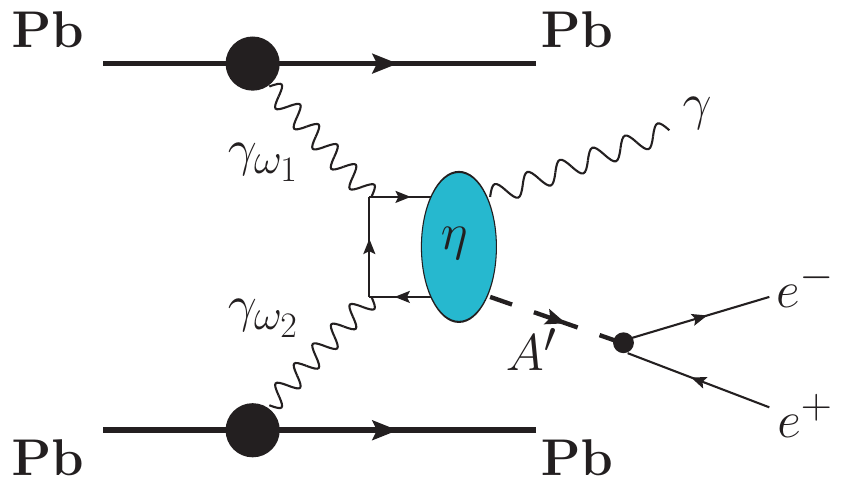}
	\caption{Feynman diagram illustrating the production of dark photons in Pb-Pb ultraperipheral collisions. The cyan shading represents the unknown details of the $\eta$ decay, and $\omega_{1(2)}$ respectively denotes the energy of the photon fluxes radiated by the two lead nuclei.}
	\label{fig:upc}
\end{figure}

The process of $PbPb$ UPC producing $\eta$ and decaying into $A^\prime$ is shown in Figure \ref{fig:upc}, where the $\eta$ meson is a photon flux generated by the interaction of heavy-ion beam radiation. UPCs require the sum of the collision parameters of the two lead nuclei to be greater than twice the nuclear radius. In UPCs, the total cross section of $PbPb\to PbPb\eta$ can be written in the well-known form form \cite{Baur:1990fx} with equivalent photon approximation \cite{Budnev:1975poe}
\begin{equation}
	\begin{aligned}
		&\sigma\left(PbPb\to Pb\otimes\eta\otimes Pb;s\right)\\
		&=\int\mathrm{d}^2\mathbf{b}_1\mathrm{d}^2\mathbf{b}_2\mathrm{d}W\mathrm{d}Y\frac W2\hat{\sigma}\left(\gamma\gamma\to\eta;W\right) \\
		&\times N\left(\omega_1,\mathbf{b}_1\right)N\left(\omega_2,\mathbf{b}_2\right)\theta(|\boldsymbol{b}_{1}-\mathbf{b}_{2}|-2R_{Pb}), 
		\end{aligned}
		\label{eq:xseciton eta}
\end{equation}
where $W=\sqrt{4\omega_1\omega_2}$ represents invariant mass of the $\gamma\gamma$ system and $Y$ is the rapidity of the $\eta$ in the final state. The photon flux energy $\omega_{1(2)}$ is written in terms of $W$ and $Y$
\begin{equation}
	\omega_1=\frac W2e^Y\quad\mathrm{and}\quad\omega_2=\frac W2e^{-Y}.
	\label{eq:WY}
\end{equation}

The cross section for the $\gamma\gamma\to\eta$ process can be calculated using
the Low formula , being given by \cite{Low:1960wv}
\begin{equation}
	\hat{\sigma}(\gamma\gamma\to \eta; W^2)=8\pi^2(2J+1)\frac{\Gamma_{\eta\to\gamma\gamma}}{m_\eta}\delta(W^2-m_\eta^2),
	\label{eq:sigmaeta}
\end{equation}
where $J=0$ is the $\eta$'s spin and $\Gamma_{\eta\to \gamma\gamma}$ is the two-photon decay width of $\eta$. Moreover,  $\boldsymbol{b}_{1}$ and $\boldsymbol{b}_{2}$ represent the impact parameters. For nucleus we choose $R_{Pb} = r_0 A_{Pb}^{1/3}$ with $r_0 = 1.2$ fm and $A_{Pb}=208$. In particular, $N(\omega,\boldsymbol{b})$ is the equivalent photon flux for a given photon energy $\omega$ and impact parameter $\boldsymbol{b}$, which can
be expressed in terms of the form factor $F(q^2)$ for the equivalent photon source as follows
\begin{equation}
	\begin{aligned}
		&N(\omega, b)=\frac{Z^{2} \alpha_{e m}}{\pi^{2}} \frac{1}{b^{2} \omega} \\
		&\times{\left[\int u^{2} J_{1}(u) F\left(\sqrt{\frac{(b \omega / \gamma_L)^{2}+u^{2}}{b^{2}}}\right) \frac{1}{(b \omega / \gamma_L)^{2}+u^{2}} \mathrm{~d} u\right]^{2} },
	\end{aligned}
	\label{eq:flux}
\end{equation}
where $Z$ is the proton number of nucleus and $\alpha_{e m}$ denotes the fine structure constant. $\gamma_L$ is the Lorentz factor which is discussed below. $J_n(u)$ is the first kind Bessel function. Form factor is the Fourier transform of charge distribution in the nucleus.
If one assume $\rho(r)$  is the spherical symmetric charge distribution, the form factor is a function
of photon virtuality $q^2$ \cite{Klusek-Gawenda:2010vqb}. 
\begin{equation}
	F\left(q^2\right)=\int \frac{4 \pi}{q} \rho(r) \sin (q r) r \mathrm{d} r=1-\frac{q^2\left\langle r^2\right\rangle}{3 !}+\frac{q^4\left\langle r^4\right\rangle}{5 !} \cdots.
	\label{eq:rr}
\end{equation}
We refer Refs. \cite{Klusek-Gawenda:2010vqb,Hencken:1995me} then the form factor function is given as 
\begin{equation}
	F(q)=\frac{\Lambda^{2}}{\Lambda^{2}+q^{2}}
	\label{eq:ffs}
\end{equation}
with $\Lambda=0.088$ GeV for nucleus \cite{Goncalves:2012cy,Goncalves:2015hra,Goncalves:2018yxc}. Thus the equivalent photon flux is written as
\begin{equation}
	\begin{aligned}
	&N(\omega,{b})=\frac{Z^2 \alpha_{\mathrm{em}}}{\pi^2} \frac{1}{\omega}\\
	&\times\left[\frac{\omega}{\gamma_L} K_1\left(b \frac{\omega}{\gamma_L}\right)-\sqrt{\frac{\omega^2}{\gamma_L^2}+\Lambda^2} K_1\left(b \sqrt{\frac{\omega^2}{\gamma_L^2}+\Lambda^2}\right)\right]^2 .
	\end{aligned}
	\label{eq:mod2}
\end{equation}
where $K_1$ is the modified Bessel function of the second kind. Lorentz factor $\gamma_L$ is computed as $\gamma_L=\sqrt{s_{NN}}/2m_N$, $m_N$ is the nucleon mass.

Based on Eqs. (\ref{eq:xseciton eta}-\ref{eq:mod2}), the total cross section of the process ${Pb}{Pb}\to Pb\otimes\gamma {e}^+{e}^-\otimes{Pb}$ is defined as 
\begin{equation}
	\begin{aligned}
	&\sigma({Pb}{Pb}\to Pb\otimes\gamma {e}^+{e}^-\otimes{Pb}; s)	=\sigma\left(PbPb\to Pb\otimes\eta\otimes Pb\right)\\
	&\times \mathcal{B}(\eta\to A^{\prime}\gamma)	  \times\mathcal{B}(A^{\prime}\to e^+e^-), 
		\end{aligned}
		\label{eq:total-xsection}
\end{equation}
where $s$ is the square of center of mas energy of PbPb system, $\mathcal{B}(\eta\to A^{\prime}\gamma) $ and $\mathcal{B}(A^{\prime}\to e^+e^-)$ denote $\eta$ decay to dark photons and dark photon decay to pairs of positive and negative electrons, respectively. In this work, we assume that $\mathcal{B}(A^{\prime}\to e^+e^-)\simeq 1$, which is valid in this analysis when the dark photon mass falls within the range of $2m_e\ll m_{A^\prime}< m_\eta$ \cite{Goncalves:2020czp}. We also find in Ref. \cite{Krovi:2019hdl} that the branching ratios for the decay of the dark photon to electron pairs and muon pairs with a mass of 0.5 GeV are both proposed to be approximately $40\%$. Additionally, the analysis of branching ratios dependent on the dark photon mass can be found in \cite{Buschmann:2015awa}. Now there is one remaining branching ratio related to the dark photon $\mathcal{B}(\eta\to A^{\prime}\gamma) $ that needs to be discussed, which is also the largest source of uncertainty in the current analysis.

\subsection{Branching ratio of $\eta \to A^\prime \gamma$}
The dark photon is the benchmark model for gauge mediators accessible at low energies \cite{Raggi:2015yfk,Fabbrichesi:2020wbt,Filippi:2020kii}. The gauge boson $A^\prime$ mixes with the photon through kinetic mixing and charge coupling \cite{okun1982limits,holdom1986two}. The kinetic mixing term is
\begin{equation}
	\mathcal{L}_{\mathrm{kin.mix.}}=-\frac\varepsilon{2\cos\theta_W}{F}_{\mu\nu}^{\prime}{B}^{\mu\nu},
	\label{eq: kine mix}
\end{equation}
where ${F}_{\mu\nu}^{\prime}({B}^{\mu\nu})$ is the U(1)$^\prime$ (hypercharge U(1)$_Y$) field strength tensor and $\theta_W$ is the weak mixing angle. This will result in a kinetic mixing coupling between the fields of the standard model and the U(1)$^\prime$ field $\mathcal{L}_{\mathrm{int}}=-e\varepsilon j_{\mathrm{em}}^\mu{A}_{\mu}^{\prime}$. Since the kinetic mixing parameter $\varepsilon$ is constrained to be small, these couplings are far weaker than electromagnetism \cite{Gan:2020aco,okun1982limits,holdom1986two,Foot:1991kb,Jaeckel:2010ni}. The couplings are suppressed by $\varepsilon$. The NA62 collaboration at the CERN SPS has reported the results of the search for $\eta$ decays to photons and dark photons $A^\prime$ \cite{NA62:2019meo}, improving the previous constraints on the dark photon mass $m_{A^\prime}$ and coupling $\epsilon^2$. However, there is currently no viable approach for the precise exploration of the parameter space of dark photons $(m_{A^\prime},\varepsilon^2)$.

Based on previous studies of the decay mode of $\eta$ into a classical photon and a dark photon, we adopt the branching ratio form from Refs. \cite{CHARM:1985nku,Pospelov:2007mp,Batell:2009di,Gninenko:2012eq,Ilten:2015hya}
\begin{equation}
	\begin{aligned}
	\mathcal{B}(\eta\to A^{\prime}\gamma)&=2\varepsilon^2\mathcal{B}(\eta\to\gamma\gamma)\left(1-m_{A^{\prime}}^2/m_\eta^2\right)^3\\
	&\times\left|\bar{{F}}_{\eta\gamma^*\gamma^*}(m_{A^{\prime}}^2,0)\right|^2,
	\end{aligned}
	\label{eq:BR}
\end{equation}
where $\mathcal{B}(\eta\to \gamma\gamma)=39.36\%$ denotes the branching ratio of $\eta$ decay to 2$\gamma$ \cite{Workman:2022ynf}. From \cite{Gninenko:2012eq,Ilten:2015hya} the transition form factor $\bar{F}_{\eta\gamma^*\gamma^*} \simeq 1$ is used in our analysis. 

We now find that in order to calculate the total cross-section Eq. (\ref{eq:total-xsection}), it is necessary to determine the values of the dark photon's mixing parameters $(m_{A^\prime},\varepsilon^2)$. As mentioned earlier, there is currently no well-established method to tightly constrain these parameters with high precision. Following suggestions from the literature review \cite{Gan:2020aco}, we limit the dynamical mixing parameter $\varepsilon$ of the dark photon to $10^{-3}< \varepsilon <10^{-2}$ \cite{Fayet:2007ua,Pospelov:2008zw}. Our choice is motivated by the significant deviation between the measurement of the $(g - 2)_\mu$ and the prediction of the standard model \cite{Muong-2:2006rrc,Aoyama:2020ynm,Muong-2:2021ojo,Muong-2:2021vma}. Introducing $A^\prime$ with the values of $\varepsilon$ within this range can reconcile this difference by BSM \cite{Gan:2020aco}.

\subsection{$PbPb$ results in UPCs }
\label{sec:UPC-results}
\begin{figure*}[htpb]
	\begin{center}
		\includegraphics[width=0.45\linewidth]{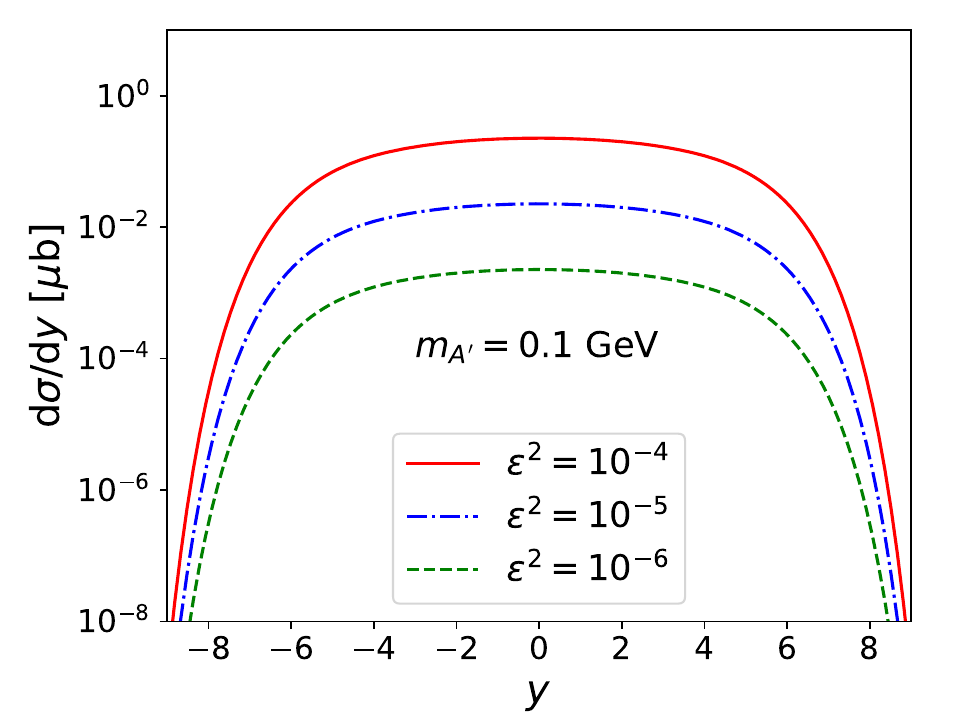}
		\includegraphics[width=0.45\linewidth]{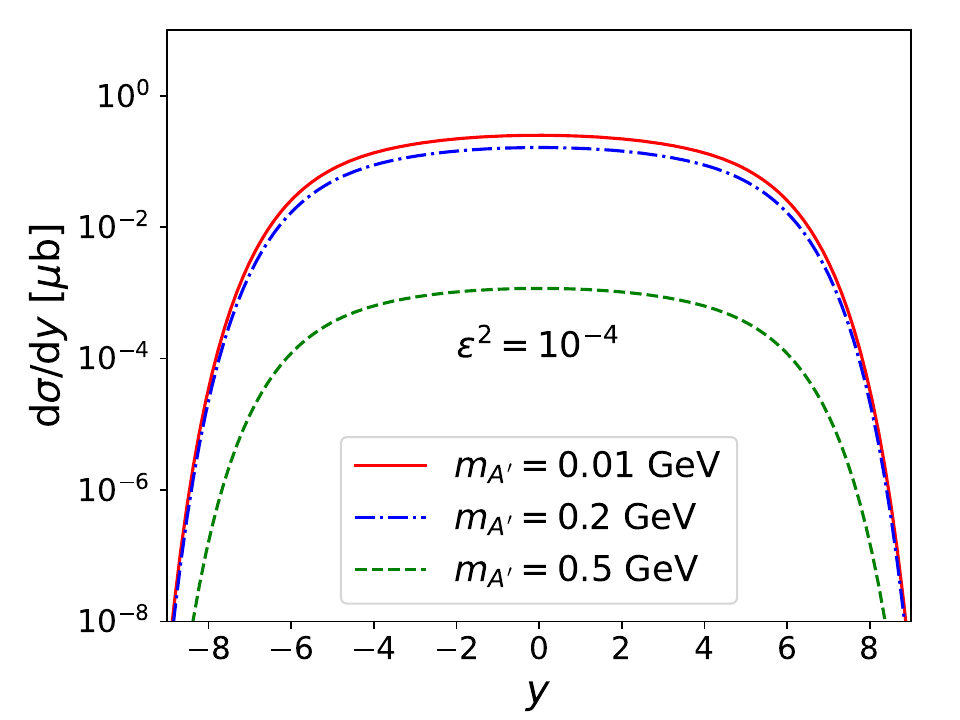}
	\end{center}
	\caption{The rapidity distribution of the differential cross-section at $\sqrt{s}=5.5$ TeV. (left) Fixing the mass of $A^\prime$ to be $m_{A^\prime}=0.1$ GeV, with $\varepsilon^2$ taking values of $10^{-4}$, $10^{-5}$, and $10^{-6}$. (right) Fixing the parameter $\varepsilon^2=10^{-4}$, with $m_{A^\prime}$ taking values of 0.01 GeV, 0.2 GeV, and 0.5 GeV.}
	\label{fig:dsigmadY}
\end{figure*}

We first show our numerical results of the total cross section of $\sigma({Pb}{Pb}\to Pb\otimes\gamma {e}^+{e}^-\otimes{Pb})$ and the rapidity distribution of differential cross section $\frac{d\sigma}{dY}$. Initially, we provide the differential cross section for $PbPb$ collisions at $\sqrt{s}=5.5$ TeV in Figure \ref{fig:dsigmadY}. We fix and differentially select the parameters associated with the two dark photons, respectively. It can be seen from Eq. (\ref{eq:BR}) that the cross-section is suppressed by the parameter $\varepsilon^2$, and the smaller the mass of the dark photon $m_{A^\prime}$, the larger the cross-section. This indicates that in the collision process under study, there is a higher likelihood of producing low-mass dark photons in the final state.

\begin{figure*}[htpb]
	\begin{center}
		\includegraphics[width=0.48\linewidth]{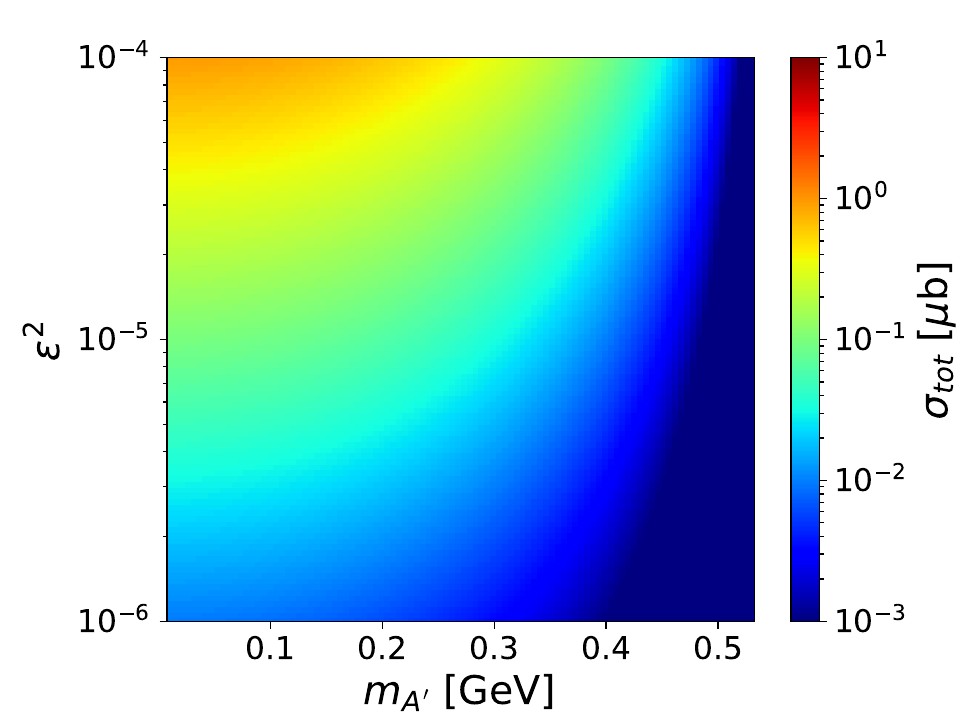}
		\includegraphics[width=0.48\linewidth]{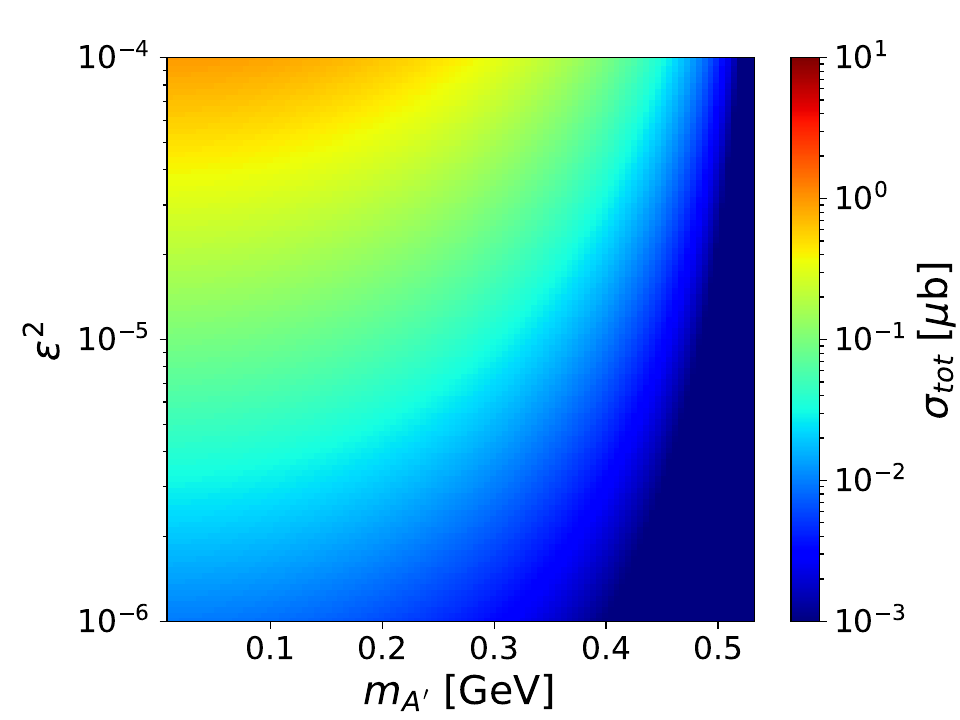}
		\includegraphics[width=0.48\linewidth]{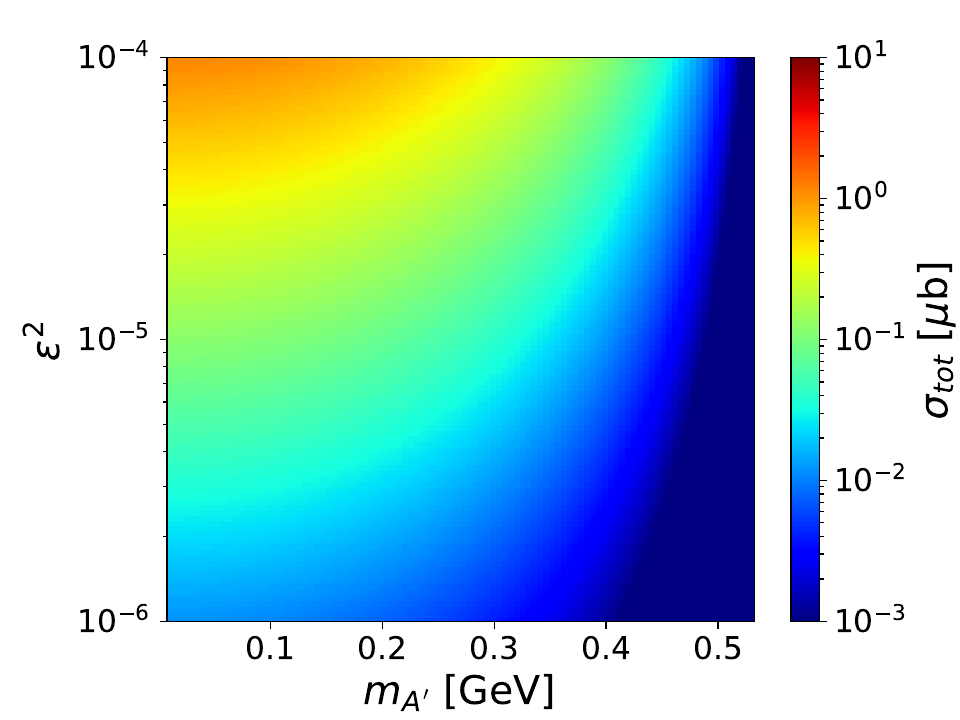}
		\includegraphics[width=0.48\linewidth]{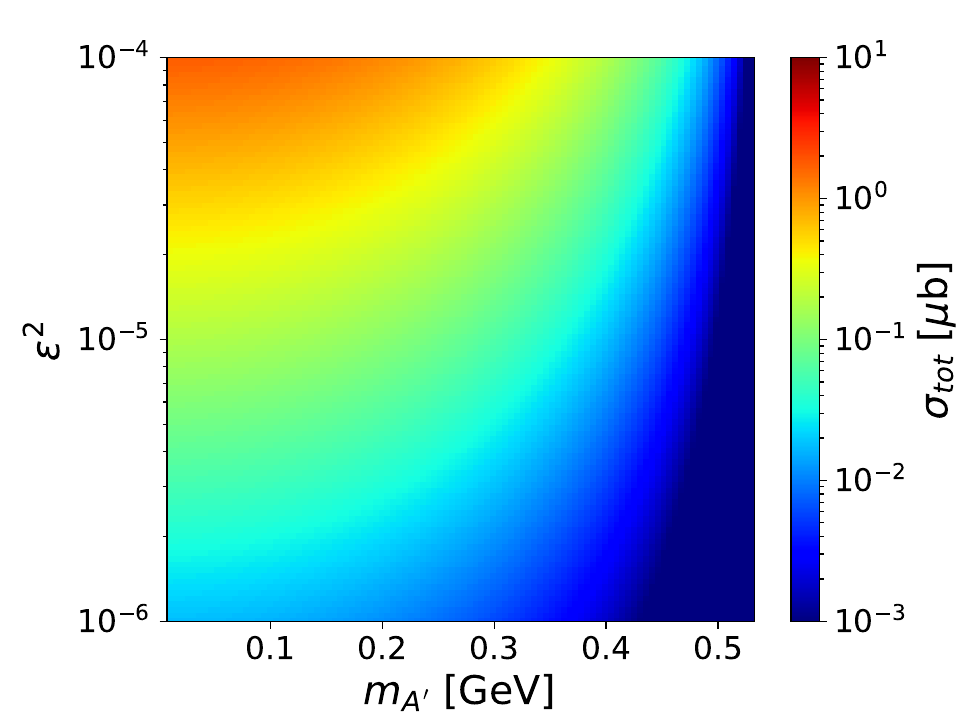}
	\end{center}
	\caption{Cross sections (in $\mu$b) for the dark photon production from $\eta$ decay in ultra-peripheral $Pb Pb$ collisions for the next run of LHC (Top left), HL-LHC (Top right), HE-LHC (Bottom left) and FCC (Bottom right).}
	\label{fig:xsection}
\end{figure*}

\begin{figure*}[htpb]
	\begin{center}
		\includegraphics[width=0.48\linewidth]{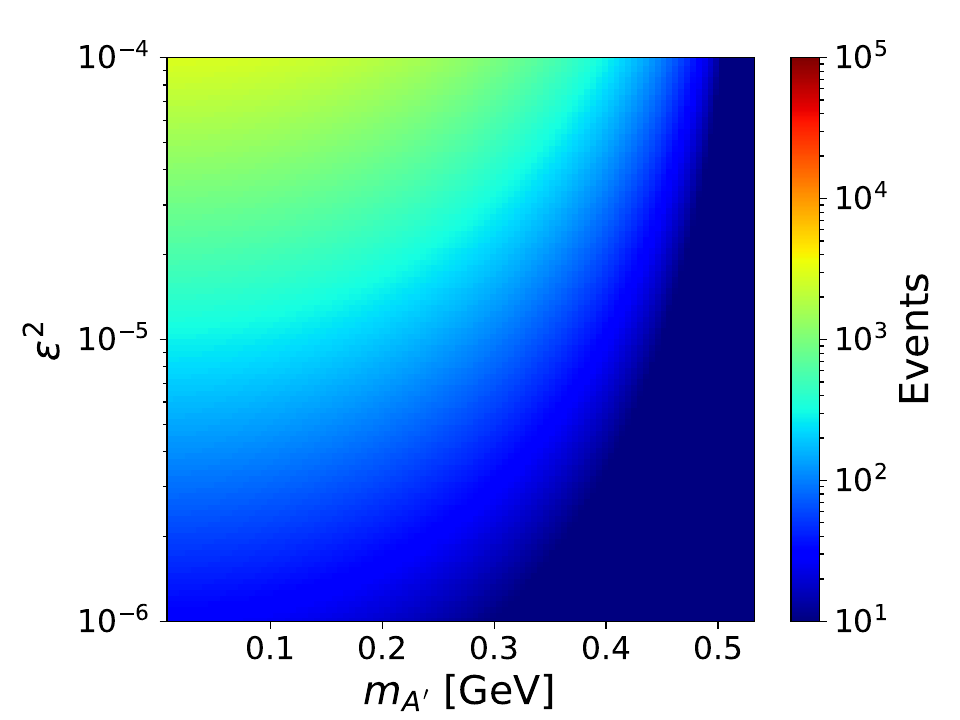}
		\includegraphics[width=0.48\linewidth]{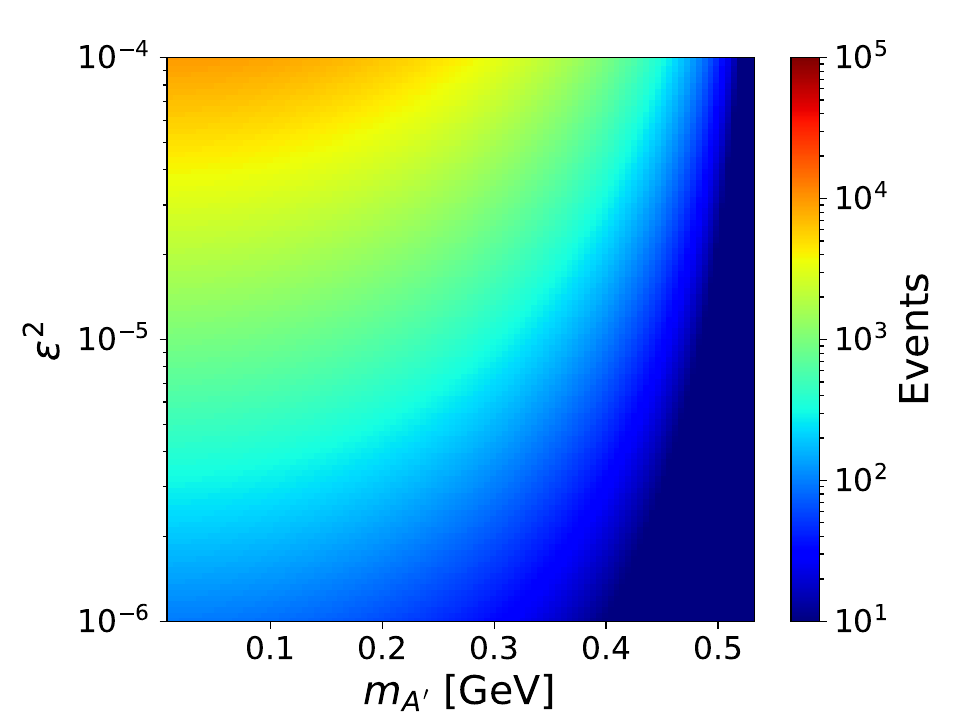}
		\includegraphics[width=0.48\linewidth]{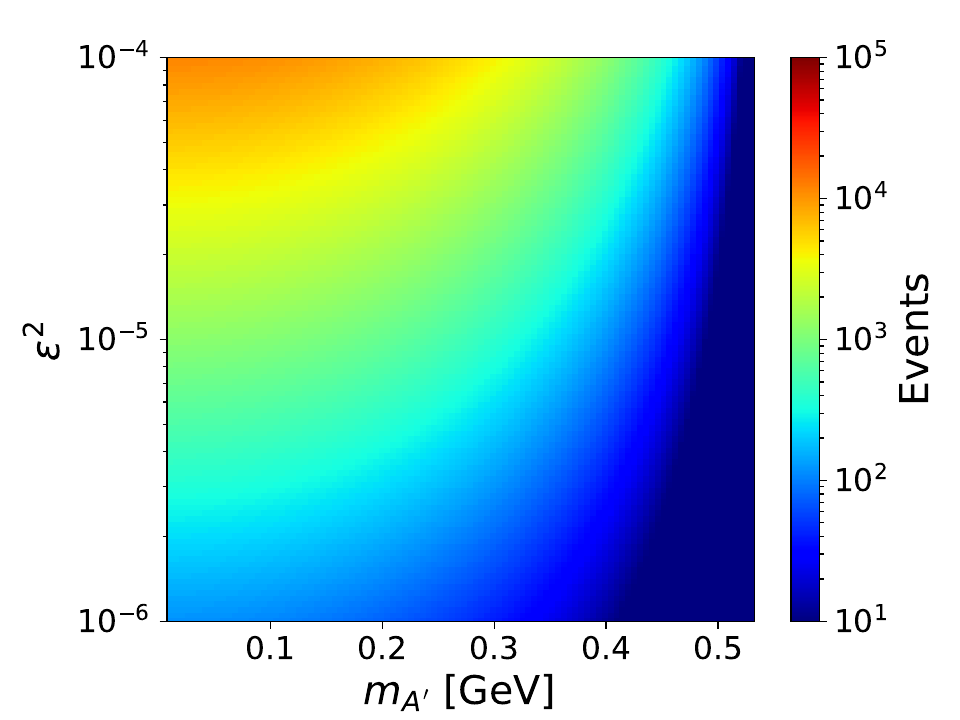}
		\includegraphics[width=0.48\linewidth]{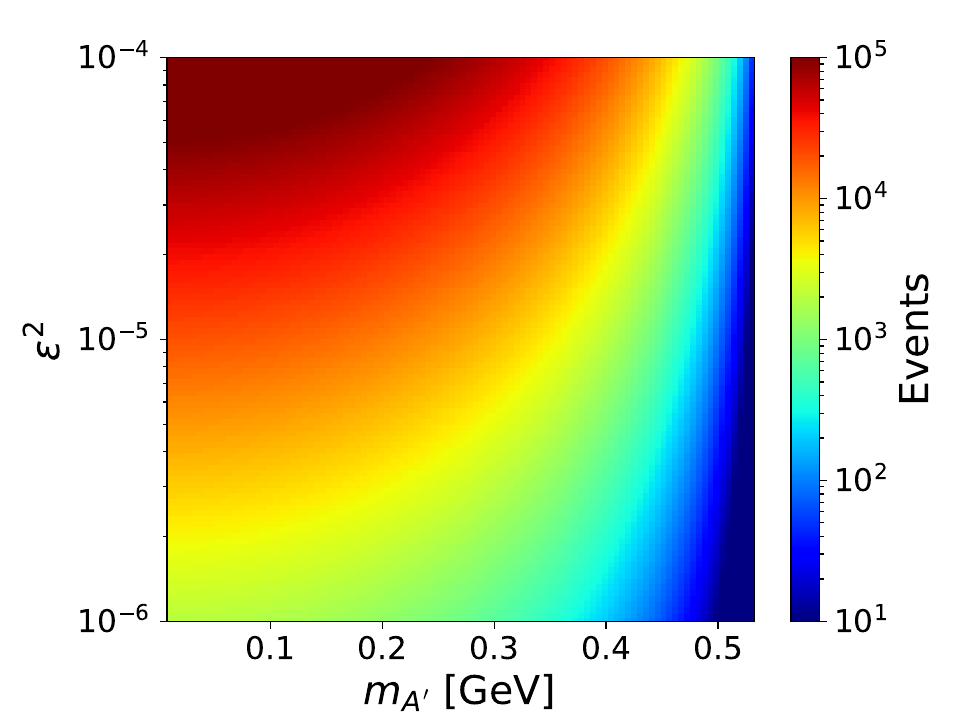}
	\end{center}
	\caption{The number of events per year for the dark photon production from $\eta$ decay in ultra-peripheral $Pb Pb$ collisions for the next run of LHC (Top left), HL-LHC (Top right), HE-LHC (Bottom left) and FCC (Bottom right).}
	\label{fig:events}
\end{figure*}

Studying the total cross-section of UPC and the number of detectable event signals in the detector is valuable. We consider the total cross-section within the detector's central rapidity interval ($|Y|<2.0$), which is a typical rapidity range for the ALICE, CMS, and ATLAS detectors at the LHC. We consider the ultra-peripheral $PbPb$ collision scenarios for the future operation and upgrade plans of the LHC. Following the references, we simultaneously estimate the total cross-section for the next run of the LHC at center-of-mass energies of $\sqrt{s}=5.5$, 5.5, 10, and 39 TeV, as well as for the High Luminosity (HL)-LHC, High Energy (HE)-LHC, and the Future Circular Collider (FCC) \cite{FCC:2018vvp,FCC:2018bvk,FCC:2018byv}. Furthermore, we assume that the luminosities for these future collider plans correspond to a full year of operation at $L=$ 3, 10, 10, and 110 nb$^{-1}$, and estimate the signal events for the production of dark photons decaying into electron-positron pairs. All the estimated results are shown in the form of a two-dimensional parameter space $(m_{A^\prime},\varepsilon^2)$ in Figures \ref{fig:xsection} and \ref{fig:events}.


Based on the above discussion, we find that the total cross-section for the production of $\eta$ decaying into dark photons in ultra-peripheral $PbPb$ collisions is suppressed by $\varepsilon^2$ and increases as the dark matter mass $m_{A^\prime}$ decreases. This result strongly depends on the decay branching ratio of $\eta\to A^\prime \gamma$  in Eq. (\ref{eq:BR}). Additionally, the total cross-section increases with the center-of-mass energy of the $PbPb$ collisions. When the parameters fall within the range $3\times 10^{-5}<\varepsilon^2<10^{-4}$ and $m_{A^\prime}<0.3$ GeV, the total cross-section is approximately greater than 0.1 $\mu$b at the four aforementioned center-of-mass energies. If we consider a set of selected parameters $\varepsilon^2=10^{-4}$, $m_{A^\prime}=50$ MeV, we can estimate $\sigma=0.95$ $\mu$b ($|Y|<2.0$). Subsequently, we briefly discuss the main background analysis, namely $\eta\to e^+e^-\gamma$, with a corresponding branching ratio of $6.9\times 10^{-3}$. Upon calculation, we find that the background cross-section is approximately $\sigma_{\mathrm{bg}}=84$ $\mu$b at $\sqrt{s}=5.5$ TeV.

Let us now analyze the estimated number of detectable events under the various device parameters mentioned above. From Figure \ref{fig:events}, it can be seen that the number of events follows the same parameter variation behavior as the total cross-section, both governed by the model (\ref{eq:BR}). Overall, a significant amount of decay events can be detected in the next run of the LHC, and the high luminosity of the heavy ion beams will increase the upper limit of observable event numbers.

\section{proton-proton collision at LHC}
In theory, in addition to heavy ion beams, proton beams can also be used for UPC experiments, although the photon flux input of UPCs is proportional to the charge number $Z$ of the beam particles, proton beams often have higher luminosity. However, higher luminosity implies larger pileup that will only make impossible this measurement in pp collisions (see the review article \cite{Soyez:2018opl}). Thus the absence of pileup in $PbPb$ collisions may make this system more beneficial to trigger on and reconstruct any UPCs compared to $pp$ collisions.

\begin{figure}[htbp]
	\centering
	\includegraphics[width=0.48\textwidth]{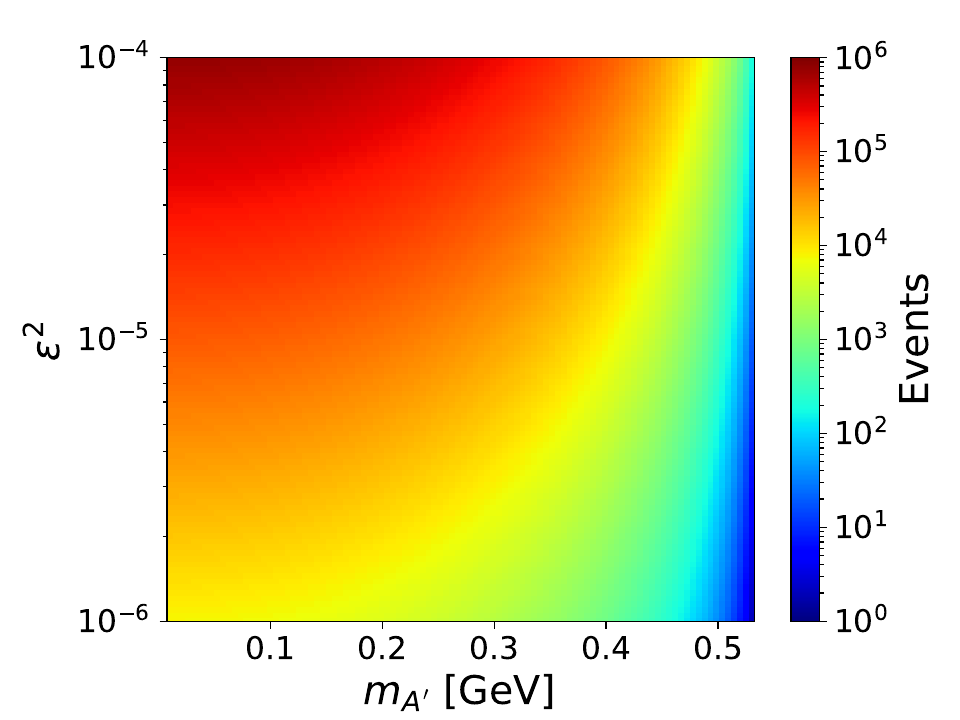}
	\caption{The estimated number of events for the production of $\eta$ decaying into dark photons in $pp$ collisions, based on the CMS Collaboration's measurement of the $\eta\to 4\mu$ branching ratio \cite{CMS:2023thf}, the uncertainties from the experimental data are not considered.}
	\label{fig:CMS}
\end{figure}

Recently, the CMS Collaboration announced their latest measurement of the branching ratio for the four-lepton decay of $\eta$ \cite{CMS:2023thf}. They analyzed a sample of proton-proton collisions collected by the CMS experiment at the CERN LHC with high-rate muon triggers during 2017 and 2018, corresponding to an integrated luminosity of 101 fb$^{-1}$. The analysis of the data yielded a branching ratio of approximately $\mathcal{B}(\eta\rightarrow\mu^+\mu^-\mu^+\mu^-)=[5.0\pm0.8(\mathrm{stat})\pm0.7(\mathrm{syst})\pm0.7(\mathcal{B}_{2\mu})]\times  10^{-9}$, with a corresponding signal yield of $N_{4\mu}=49.6\pm8.1$. We utilize this result to estimate how many dark photon events are produced by the proton-proton collision data they used to generate the $\eta$ mesons. To simplify the analysis, we assume a uniform detector acceptance. We estimate the number of events for $A^\prime$ production from $\eta$ decays by the form 
\begin{equation}
	\begin{aligned}
		&N_{4\mu}=\mathcal{L}_{int} \times \sigma_{pp\to \eta}\times \mathcal{B}(\eta\to4\mu)\times\beta,\\
		&N_{A^\prime\gamma}=\mathcal{L}_{int} \times \sigma_{pp\to \eta}\times \mathcal{B}(\eta\to A^\prime \gamma)\times\beta,
		\end{aligned}
			\label{eq:CMS-estimate}
\end{equation}
where $\mathcal{L}_{int}$ is the integrated luminosity at LHC we considered. The detector acceptance $\beta$ is same in both equations in Eq. (\ref{eq:CMS-estimate}). Based on Eqs. (\ref{eq:BR}-\ref{eq:CMS-estimate}) we thus have
\begin{equation}
	N_{A^\prime\gamma}=\frac{\mathcal{B}(\eta\to 2\gamma)}{\mathcal{B}(\eta\to4\mu)} 2\varepsilon^2\left(1-\frac{m_{A^\prime}^2}{m_\eta^2}\right)^3\left|\bar{{F}}_{\eta\gamma^*\gamma^*}(m_{A^{\prime}}^2,0)\right|^2.
	\label{eq:events of LHC}
\end{equation}
The estimation of events at CMS by LHC is shown in Figure \ref{fig:CMS}.
One can see that the distribution of dark photon event estimates in the $(m_{A^\prime},\varepsilon^2)$ plane, based on the branching ratio analysis provided by the CMS Collaboration, is significant. The highest signal yield can reach around $10^6$ within the parameter range we provided. We believe that searching for dark photons in the current LHC $pp$ collision experiments is promising.

We also need to clarify the numerical results above. Firstly, we make the strong assumption that the dark photon decays into a positron-electron pair with a branching ratio of $1.0$. This is a strong assumption, as the decay behavior of dark photons is currently not well-explored experimentally. Nevertheless, considering discussions similar to those in Ref. \cite{Gninenko:2012eq} regarding $A^\prime\to e^+e^-$ is valuable. Secondly, there are other related studies on meson decays into dark photons, and the branching ratio we use is an approximate result, which does not account for the impact of the actual transition form factor of $\eta$ to double virtual photons. For further investigation on this aspect, please refer to Ref. \cite{Choi:2019wqx}, which is based on a light-front quark model. Lastly, we explore the recent CMS analysis as a benchmark for the highest number of $\eta$ produced in proton-proton collisions and estimate the number of dark photon events based on this. It can be further extended to other leptonic decay channels of the dark photon, such as $\mu^+\mu^-$. Our results may provide some positive insights for the search for dark photons.

\section{Summary and outlook}
\label{sec:summary}
In this study, we investigate the production of dark photons from the decay of $\eta$ mesons in UPC at the LHC. We utilize the $\eta\to A^\prime \gamma$ branching ratio descriptions from Refs. \cite{CHARM:1985nku,Pospelov:2007mp,Batell:2009di,Gninenko:2012eq,Ilten:2015hya}, and assume the branching ratio of $A^\prime$ decaying into positron-electron pairs. By performing numerical calculations, we estimate the total cross-section for producing dark photons in ultra-peripheral $PbPb$ collisions at existing and planned future hadron colliders, as well as the signal yields, as depicted in Figures \ref{fig:xsection} and \ref{fig:events}. We also briefly discuss the uncertainties that may arise in the calculation process. Overall, the kinematic mixing parameter $\varepsilon^2$ and the dark photon mass $m_{A^\prime}$ primarily govern the entire process.

Additionally, we have conducted a preliminary assessment of the proposed $\eta$ factory and believe that it will significantly increase the probability of finding dark photons. We then utilize the recent measurement by the CMS Collaboration of the $\eta\to 4\mu$ branching ratio to estimate the number of $\eta$ decays into dark photons (see Figure \ref{fig:CMS}). The results indicate that it is very likely to find dark photons in the total $\eta$ decay events. We propose to search for dark photons $A^\prime$ in the $\eta$ decay processes produced in the proton-proton collision experiments at the existing LHC. Specifically, reconstructing the number of events of dark photons in the decay final states of leptonic pairs is timely, such as $e^+e^-$ for ATLAS and $\mu^+\mu^-$ for CMS ($m_{A^\prime}>2 m_\mu$). This is because the performance of the CMS detector in measuring low-transverse momentum electrons is unsatisfactory, and there are also some deficiencies in particle identification in the calorimeter. Therefore, it may be worth considering reconstructing the $e^+e^-\gamma$ final state on ATLAS to constrain the number of dark photon events.

It should be noted that high-precision measurements of the $\eta$ decay channel often require a higher statistical yield of $\eta$ mesons. In addition to existing heavy-ion colliders and the existing electron-ion collider at the Jefferson Lab, planned future $\eta$ factories could significantly increase the yield of $\eta$ mesons. A new experiment, REDTOP (Rare Eta Decays To Probe New Physics) \cite{REDTOP:2022slw}, is being proposed, with the intent of collecting a data sample of the order of $10^{14}\ \eta$ (10$^{12}\ \eta^\prime$) for studying very rare decays. Moreover, the Primakoff process \cite{Primakoff:1951iae} at high-energy electron-ion colliders in the United States  \cite{Accardi:2012qut,AbdulKhalek:2021gbh} and China \cite{Chen:2018wyz,Chen:2020ijn,Anderle:2021wcy} can also yield a significant number of $\eta$ mesons. In addition to the common two-photon decay channel, the measurement precision of rare $\eta$ decays can also be significantly improved.

\begin{acknowledgments}
 We are very grateful for the valuable discussion and communication about CMS experiment with Hao Qiu and Shuai Yang. This work is supported by the Strategic Priority Research Program of Chinese Academy of Sciences under the Grant NO. XDB34030301.
%	and
%	Guangdong Major Project of Basic and Applied Basic
%	Research NO. 2020B0301030008.
\end{acknowledgments}

\bibliographystyle{apsrev4-1}
\bibliography{refs}

\begin{thebibliography}{71}%
\makeatletter
\providecommand \@ifxundefined [1]{%
 \@ifx{#1\undefined}
}%
\providecommand \@ifnum [1]{%
 \ifnum #1\expandafter \@firstoftwo
 \else \expandafter \@secondoftwo
 \fi
}%
\providecommand \@ifx [1]{%
 \ifx #1\expandafter \@firstoftwo
 \else \expandafter \@secondoftwo
 \fi
}%
\providecommand \natexlab [1]{#1}%
\providecommand \enquote  [1]{``#1''}%
\providecommand \bibnamefont  [1]{#1}%
\providecommand \bibfnamefont [1]{#1}%
\providecommand \citenamefont [1]{#1}%
\providecommand \href@noop [0]{\@secondoftwo}%
\providecommand \href [0]{\begingroup \@sanitize@url \@href}%
\providecommand \@href[1]{\@@startlink{#1}\@@href}%
\providecommand \@@href[1]{\endgroup#1\@@endlink}%
\providecommand \@sanitize@url [0]{\catcode `\\12\catcode `\$12\catcode
  `\&12\catcode `\#12\catcode `\^12\catcode `\_12\catcode `\%12\relax}%
\providecommand \@@startlink[1]{}%
\providecommand \@@endlink[0]{}%
\providecommand \url  [0]{\begingroup\@sanitize@url \@url }%
\providecommand \@url [1]{\endgroup\@href {#1}{\urlprefix }}%
\providecommand \urlprefix  [0]{URL }%
\providecommand \Eprint [0]{\href }%
\providecommand \doibase [0]{http://dx.doi.org/}%
\providecommand \selectlanguage [0]{\@gobble}%
\providecommand \bibinfo  [0]{\@secondoftwo}%
\providecommand \bibfield  [0]{\@secondoftwo}%
\providecommand \translation [1]{[#1]}%
\providecommand \BibitemOpen [0]{}%
\providecommand \bibitemStop [0]{}%
\providecommand \bibitemNoStop [0]{.\EOS\space}%
\providecommand \EOS [0]{\spacefactor3000\relax}%
\providecommand \BibitemShut  [1]{\csname bibitem#1\endcsname}%
\let\auto@bib@innerbib\@empty
%</preamble>
\bibitem [{\citenamefont {Sakata}\ \emph {et~al.}(1956)\citenamefont {Sakata}
  \emph {et~al.}}]{sakata1956composite}%
  \BibitemOpen
  \bibfield  {author} {\bibinfo {author} {\bibfnamefont {S.}~\bibnamefont
  {Sakata}} \emph {et~al.},\ }\href@noop {} {\bibfield  {journal} {\bibinfo
  {journal} {Prog. Theor. Phys}\ }\textbf {\bibinfo {volume} {16}},\ \bibinfo
  {pages} {686} (\bibinfo {year} {1956})}\BibitemShut {NoStop}%
\bibitem [{\citenamefont {Okun}(1958)}]{okun1958some}%
  \BibitemOpen
  \bibfield  {author} {\bibinfo {author} {\bibfnamefont {L.}~\bibnamefont
  {Okun}},\ }\href@noop {} {\bibfield  {journal} {\bibinfo  {journal} {Zhur.
  Eksptl'. i Teoret. Fiz.}\ }\textbf {\bibinfo {volume} {34}} (\bibinfo {year}
  {1958})}\BibitemShut {NoStop}%
\bibitem [{\citenamefont {Yamaguchi}(1958)}]{yamaguchi1958possible}%
  \BibitemOpen
  \bibfield  {author} {\bibinfo {author} {\bibfnamefont {Y.}~\bibnamefont
  {Yamaguchi}},\ }\href@noop {} {\bibfield  {journal} {\bibinfo  {journal}
  {Progress of Theoretical Physics}\ }\textbf {\bibinfo {volume} {19}}
  (\bibinfo {year} {1958})}\BibitemShut {NoStop}%
\bibitem [{\citenamefont {Yamaguchi}(1959)}]{yamaguchi1959model}%
  \BibitemOpen
  \bibfield  {author} {\bibinfo {author} {\bibfnamefont {Y.}~\bibnamefont
  {Yamaguchi}},\ }\href@noop {} {\bibfield  {journal} {\bibinfo  {journal}
  {Progress of Theoretical Physics Supplement}\ }\textbf {\bibinfo {volume}
  {11}},\ \bibinfo {pages} {37} (\bibinfo {year} {1959})}\BibitemShut {NoStop}%
\bibitem [{\citenamefont {Ikeda}\ \emph {et~al.}(1959)\citenamefont {Ikeda},
  \citenamefont {Ogawa},\ and\ \citenamefont {Ohnuki}}]{ikeda1959possibile}%
  \BibitemOpen
  \bibfield  {author} {\bibinfo {author} {\bibfnamefont {M.}~\bibnamefont
  {Ikeda}}, \bibinfo {author} {\bibfnamefont {S.}~\bibnamefont {Ogawa}}, \ and\
  \bibinfo {author} {\bibfnamefont {Y.}~\bibnamefont {Ohnuki}},\ }\href@noop {}
  {\bibfield  {journal} {\bibinfo  {journal} {Progress of Theoretical Physics}\
  }\textbf {\bibinfo {volume} {22}},\ \bibinfo {pages} {715} (\bibinfo {year}
  {1959})}\BibitemShut {NoStop}%
\bibitem [{\citenamefont {Gell-Mann}(1961)}]{Gell-Mann:1961omu}%
  \BibitemOpen
  \bibfield  {author} {\bibinfo {author} {\bibfnamefont {M.}~\bibnamefont
  {Gell-Mann}},\ }\href {\doibase 10.2172/4008239} {\  (\bibinfo {year}
  {1961}),\ 10.2172/4008239}\BibitemShut {NoStop}%
\bibitem [{\citenamefont {Ne'eman}(1961)}]{ne1961derivation}%
  \BibitemOpen
  \bibfield  {author} {\bibinfo {author} {\bibfnamefont {Y.}~\bibnamefont
  {Ne'eman}},\ }\href@noop {} {\bibfield  {journal} {\bibinfo  {journal}
  {Nuclear physics}\ }\textbf {\bibinfo {volume} {26}},\ \bibinfo {pages} {222}
  (\bibinfo {year} {1961})}\BibitemShut {NoStop}%
\bibitem [{\citenamefont {Pevsner}\ \emph {et~al.}(1961)\citenamefont {Pevsner}
  \emph {et~al.}}]{Pevsner:1961pa}%
  \BibitemOpen
  \bibfield  {author} {\bibinfo {author} {\bibfnamefont {A.}~\bibnamefont
  {Pevsner}} \emph {et~al.},\ }\href {\doibase 10.1103/PhysRevLett.7.421}
  {\bibfield  {journal} {\bibinfo  {journal} {Phys. Rev. Lett.}\ }\textbf
  {\bibinfo {volume} {7}},\ \bibinfo {pages} {421} (\bibinfo {year}
  {1961})}\BibitemShut {NoStop}%
\bibitem [{\citenamefont {Gan}\ \emph {et~al.}(2022)\citenamefont {Gan},
  \citenamefont {Kubis}, \citenamefont {Passemar},\ and\ \citenamefont
  {Tulin}}]{Gan:2020aco}%
  \BibitemOpen
  \bibfield  {author} {\bibinfo {author} {\bibfnamefont {L.}~\bibnamefont
  {Gan}}, \bibinfo {author} {\bibfnamefont {B.}~\bibnamefont {Kubis}}, \bibinfo
  {author} {\bibfnamefont {E.}~\bibnamefont {Passemar}}, \ and\ \bibinfo
  {author} {\bibfnamefont {S.}~\bibnamefont {Tulin}},\ }\href {\doibase
  10.1016/j.physrep.2021.11.001} {\bibfield  {journal} {\bibinfo  {journal}
  {Phys. Rept.}\ }\textbf {\bibinfo {volume} {945}},\ \bibinfo {pages} {1}
  (\bibinfo {year} {2022})},\ \Eprint {http://arxiv.org/abs/2007.00664}
  {arXiv:2007.00664 [hep-ph]} \BibitemShut {NoStop}%
\bibitem [{\citenamefont {Essig}\ \emph {et~al.}(2013)\citenamefont {Essig}
  \emph {et~al.}}]{Essig:2013lka}%
  \BibitemOpen
  \bibfield  {author} {\bibinfo {author} {\bibfnamefont {R.}~\bibnamefont
  {Essig}} \emph {et~al.},\ }in\ \href@noop {} {\emph {\bibinfo {booktitle}
  {{Snowmass 2013}: {Snowmass on the Mississippi}}}}\ (\bibinfo {year} {2013})\
  \Eprint {http://arxiv.org/abs/1311.0029} {arXiv:1311.0029 [hep-ph]}
  \BibitemShut {NoStop}%
\bibitem [{\citenamefont {Alekhin}\ \emph {et~al.}(2016)\citenamefont {Alekhin}
  \emph {et~al.}}]{Alekhin:2015byh}%
  \BibitemOpen
  \bibfield  {author} {\bibinfo {author} {\bibfnamefont {S.}~\bibnamefont
  {Alekhin}} \emph {et~al.},\ }\href {\doibase 10.1088/0034-4885/79/12/124201}
  {\bibfield  {journal} {\bibinfo  {journal} {Rept. Prog. Phys.}\ }\textbf
  {\bibinfo {volume} {79}},\ \bibinfo {pages} {124201} (\bibinfo {year}
  {2016})},\ \Eprint {http://arxiv.org/abs/1504.04855} {arXiv:1504.04855
  [hep-ph]} \BibitemShut {NoStop}%
\bibitem [{\citenamefont {Battaglieri}\ \emph {et~al.}(2017)\citenamefont
  {Battaglieri} \emph {et~al.}}]{Battaglieri:2017aum}%
  \BibitemOpen
  \bibfield  {author} {\bibinfo {author} {\bibfnamefont {M.}~\bibnamefont
  {Battaglieri}} \emph {et~al.},\ }in\ \href@noop {} {\emph {\bibinfo
  {booktitle} {{U.S. Cosmic Visions: New Ideas in Dark Matter}}}}\ (\bibinfo
  {year} {2017})\ \Eprint {http://arxiv.org/abs/1707.04591} {arXiv:1707.04591
  [hep-ph]} \BibitemShut {NoStop}%
\bibitem [{\citenamefont {Alexander}\ \emph {et~al.}(2016)\citenamefont
  {Alexander} \emph {et~al.}}]{Alexander:2016aln}%
  \BibitemOpen
  \bibfield  {author} {\bibinfo {author} {\bibfnamefont {J.}~\bibnamefont
  {Alexander}} \emph {et~al.}\ }(\bibinfo {year} {2016})\ \Eprint
  {http://arxiv.org/abs/1608.08632} {arXiv:1608.08632 [hep-ph]} \BibitemShut
  {NoStop}%
\bibitem [{\citenamefont {Beacham}\ \emph {et~al.}(2020)\citenamefont {Beacham}
  \emph {et~al.}}]{Beacham:2019nyx}%
  \BibitemOpen
  \bibfield  {author} {\bibinfo {author} {\bibfnamefont {J.}~\bibnamefont
  {Beacham}} \emph {et~al.},\ }\href {\doibase 10.1088/1361-6471/ab4cd2}
  {\bibfield  {journal} {\bibinfo  {journal} {J. Phys. G}\ }\textbf {\bibinfo
  {volume} {47}},\ \bibinfo {pages} {010501} (\bibinfo {year} {2020})},\
  \Eprint {http://arxiv.org/abs/1901.09966} {arXiv:1901.09966 [hep-ex]}
  \BibitemShut {NoStop}%
\bibitem [{\citenamefont {Nelson}\ and\ \citenamefont
  {Tetradis}(1989)}]{Nelson:1989fx}%
  \BibitemOpen
  \bibfield  {author} {\bibinfo {author} {\bibfnamefont {A.~E.}\ \bibnamefont
  {Nelson}}\ and\ \bibinfo {author} {\bibfnamefont {N.}~\bibnamefont
  {Tetradis}},\ }\href {\doibase 10.1016/0370-2693(89)90196-2} {\bibfield
  {journal} {\bibinfo  {journal} {Phys. Lett. B}\ }\textbf {\bibinfo {volume}
  {221}},\ \bibinfo {pages} {80} (\bibinfo {year} {1989})}\BibitemShut
  {NoStop}%
\bibitem [{\citenamefont {Fayet}(2006)}]{Fayet:2006sp}%
  \BibitemOpen
  \bibfield  {author} {\bibinfo {author} {\bibfnamefont {P.}~\bibnamefont
  {Fayet}},\ }\href {\doibase 10.1103/PhysRevD.74.054034} {\bibfield  {journal}
  {\bibinfo  {journal} {Phys. Rev. D}\ }\textbf {\bibinfo {volume} {74}},\
  \bibinfo {pages} {054034} (\bibinfo {year} {2006})},\ \Eprint
  {http://arxiv.org/abs/hep-ph/0607318} {arXiv:hep-ph/0607318} \BibitemShut
  {NoStop}%
\bibitem [{\citenamefont {Boveia}\ and\ \citenamefont
  {Doglioni}(2018)}]{Boveia:2018yeb}%
  \BibitemOpen
  \bibfield  {author} {\bibinfo {author} {\bibfnamefont {A.}~\bibnamefont
  {Boveia}}\ and\ \bibinfo {author} {\bibfnamefont {C.}~\bibnamefont
  {Doglioni}},\ }\href {\doibase 10.1146/annurev-nucl-101917-021008} {\bibfield
   {journal} {\bibinfo  {journal} {Ann. Rev. Nucl. Part. Sci.}\ }\textbf
  {\bibinfo {volume} {68}},\ \bibinfo {pages} {429} (\bibinfo {year} {2018})},\
  \Eprint {http://arxiv.org/abs/1810.12238} {arXiv:1810.12238 [hep-ex]}
  \BibitemShut {NoStop}%
\bibitem [{\citenamefont {Filippi}\ and\ \citenamefont
  {De~Napoli}(2020)}]{Filippi:2020kii}%
  \BibitemOpen
  \bibfield  {author} {\bibinfo {author} {\bibfnamefont {A.}~\bibnamefont
  {Filippi}}\ and\ \bibinfo {author} {\bibfnamefont {M.}~\bibnamefont
  {De~Napoli}},\ }\href {\doibase 10.1016/j.revip.2020.100042} {\bibfield
  {journal} {\bibinfo  {journal} {Rev. Phys.}\ }\textbf {\bibinfo {volume}
  {5}},\ \bibinfo {pages} {100042} (\bibinfo {year} {2020})},\ \Eprint
  {http://arxiv.org/abs/2006.04640} {arXiv:2006.04640 [hep-ph]} \BibitemShut
  {NoStop}%
\bibitem [{\citenamefont {Batley}\ \emph {et~al.}(2015)\citenamefont {Batley}
  \emph {et~al.}}]{NA482:2015wmo}%
  \BibitemOpen
  \bibfield  {author} {\bibinfo {author} {\bibfnamefont {J.~R.}\ \bibnamefont
  {Batley}} \emph {et~al.} (\bibinfo {collaboration} {NA48/2}),\ }\href
  {\doibase 10.1016/j.physletb.2015.04.068} {\bibfield  {journal} {\bibinfo
  {journal} {Phys. Lett. B}\ }\textbf {\bibinfo {volume} {746}},\ \bibinfo
  {pages} {178} (\bibinfo {year} {2015})},\ \Eprint
  {http://arxiv.org/abs/1504.00607} {arXiv:1504.00607 [hep-ex]} \BibitemShut
  {NoStop}%
\bibitem [{\citenamefont {Anastasi}\ \emph {et~al.}(2016)\citenamefont
  {Anastasi} \emph {et~al.}}]{KLOE-2:2016ydq}%
  \BibitemOpen
  \bibfield  {author} {\bibinfo {author} {\bibfnamefont {A.}~\bibnamefont
  {Anastasi}} \emph {et~al.} (\bibinfo {collaboration} {KLOE-2}),\ }\href
  {\doibase 10.1016/j.physletb.2016.04.019} {\bibfield  {journal} {\bibinfo
  {journal} {Phys. Lett. B}\ }\textbf {\bibinfo {volume} {757}},\ \bibinfo
  {pages} {356} (\bibinfo {year} {2016})},\ \Eprint
  {http://arxiv.org/abs/1603.06086} {arXiv:1603.06086 [hep-ex]} \BibitemShut
  {NoStop}%
\bibitem [{\citenamefont {Lees}\ \emph {et~al.}(2017)\citenamefont {Lees} \emph
  {et~al.}}]{BaBar:2017tiz}%
  \BibitemOpen
  \bibfield  {author} {\bibinfo {author} {\bibfnamefont {J.~P.}\ \bibnamefont
  {Lees}} \emph {et~al.} (\bibinfo {collaboration} {BaBar}),\ }\href {\doibase
  10.1103/PhysRevLett.119.131804} {\bibfield  {journal} {\bibinfo  {journal}
  {Phys. Rev. Lett.}\ }\textbf {\bibinfo {volume} {119}},\ \bibinfo {pages}
  {131804} (\bibinfo {year} {2017})},\ \Eprint
  {http://arxiv.org/abs/1702.03327} {arXiv:1702.03327 [hep-ex]} \BibitemShut
  {NoStop}%
\bibitem [{\citenamefont {Aaij}\ \emph {et~al.}(2018)\citenamefont {Aaij} \emph
  {et~al.}}]{LHCb:2017trq}%
  \BibitemOpen
  \bibfield  {author} {\bibinfo {author} {\bibfnamefont {R.}~\bibnamefont
  {Aaij}} \emph {et~al.} (\bibinfo {collaboration} {LHCb}),\ }\href {\doibase
  10.1103/PhysRevLett.120.061801} {\bibfield  {journal} {\bibinfo  {journal}
  {Phys. Rev. Lett.}\ }\textbf {\bibinfo {volume} {120}},\ \bibinfo {pages}
  {061801} (\bibinfo {year} {2018})},\ \Eprint
  {http://arxiv.org/abs/1710.02867} {arXiv:1710.02867 [hep-ex]} \BibitemShut
  {NoStop}%
\bibitem [{\citenamefont {Adrian}\ \emph {et~al.}(2018)\citenamefont {Adrian}
  \emph {et~al.}}]{HPS:2018xkw}%
  \BibitemOpen
  \bibfield  {author} {\bibinfo {author} {\bibfnamefont {P.~H.}\ \bibnamefont
  {Adrian}} \emph {et~al.} (\bibinfo {collaboration} {HPS}),\ }\href {\doibase
  10.1103/PhysRevD.98.091101} {\bibfield  {journal} {\bibinfo  {journal} {Phys.
  Rev. D}\ }\textbf {\bibinfo {volume} {98}},\ \bibinfo {pages} {091101}
  (\bibinfo {year} {2018})},\ \Eprint {http://arxiv.org/abs/1807.11530}
  {arXiv:1807.11530 [hep-ex]} \BibitemShut {NoStop}%
\bibitem [{\citenamefont {Aaij}\ \emph {et~al.}(2020)\citenamefont {Aaij} \emph
  {et~al.}}]{LHCb:2019vmc}%
  \BibitemOpen
  \bibfield  {author} {\bibinfo {author} {\bibfnamefont {R.}~\bibnamefont
  {Aaij}} \emph {et~al.} (\bibinfo {collaboration} {LHCb}),\ }\href {\doibase
  10.1103/PhysRevLett.124.041801} {\bibfield  {journal} {\bibinfo  {journal}
  {Phys. Rev. Lett.}\ }\textbf {\bibinfo {volume} {124}},\ \bibinfo {pages}
  {041801} (\bibinfo {year} {2020})},\ \Eprint
  {http://arxiv.org/abs/1910.06926} {arXiv:1910.06926 [hep-ex]} \BibitemShut
  {NoStop}%
\bibitem [{\citenamefont {Cortina~Gil}\ \emph {et~al.}(2019)\citenamefont
  {Cortina~Gil} \emph {et~al.}}]{NA62:2019meo}%
  \BibitemOpen
  \bibfield  {author} {\bibinfo {author} {\bibfnamefont {E.}~\bibnamefont
  {Cortina~Gil}} \emph {et~al.} (\bibinfo {collaboration} {NA62}),\ }\href
  {\doibase 10.1007/JHEP05(2019)182} {\bibfield  {journal} {\bibinfo  {journal}
  {JHEP}\ }\textbf {\bibinfo {volume} {05}},\ \bibinfo {pages} {182} (\bibinfo
  {year} {2019})},\ \Eprint {http://arxiv.org/abs/1903.08767} {arXiv:1903.08767
  [hep-ex]} \BibitemShut {NoStop}%
\bibitem [{\citenamefont {Goncalves}\ and\ \citenamefont
  {Moreira}(2020)}]{Goncalves:2020czp}%
  \BibitemOpen
  \bibfield  {author} {\bibinfo {author} {\bibfnamefont {V.~P.}\ \bibnamefont
  {Goncalves}}\ and\ \bibinfo {author} {\bibfnamefont {B.~D.}\ \bibnamefont
  {Moreira}},\ }\href {\doibase 10.1016/j.physletb.2020.135635} {\bibfield
  {journal} {\bibinfo  {journal} {Phys. Lett. B}\ }\textbf {\bibinfo {volume}
  {808}},\ \bibinfo {pages} {135635} (\bibinfo {year} {2020})},\ \Eprint
  {http://arxiv.org/abs/2006.08348} {arXiv:2006.08348 [hep-ph]} \BibitemShut
  {NoStop}%
\bibitem [{\citenamefont {Batell}\ \emph {et~al.}(2009)\citenamefont {Batell},
  \citenamefont {Pospelov},\ and\ \citenamefont {Ritz}}]{Batell:2009di}%
  \BibitemOpen
  \bibfield  {author} {\bibinfo {author} {\bibfnamefont {B.}~\bibnamefont
  {Batell}}, \bibinfo {author} {\bibfnamefont {M.}~\bibnamefont {Pospelov}}, \
  and\ \bibinfo {author} {\bibfnamefont {A.}~\bibnamefont {Ritz}},\ }\href
  {\doibase 10.1103/PhysRevD.80.095024} {\bibfield  {journal} {\bibinfo
  {journal} {Phys. Rev. D}\ }\textbf {\bibinfo {volume} {80}},\ \bibinfo
  {pages} {095024} (\bibinfo {year} {2009})},\ \Eprint
  {http://arxiv.org/abs/0906.5614} {arXiv:0906.5614 [hep-ph]} \BibitemShut
  {NoStop}%
\bibitem [{\citenamefont {deNiverville}\ \emph {et~al.}(2011)\citenamefont
  {deNiverville}, \citenamefont {Pospelov},\ and\ \citenamefont
  {Ritz}}]{deNiverville:2011it}%
  \BibitemOpen
  \bibfield  {author} {\bibinfo {author} {\bibfnamefont {P.}~\bibnamefont
  {deNiverville}}, \bibinfo {author} {\bibfnamefont {M.}~\bibnamefont
  {Pospelov}}, \ and\ \bibinfo {author} {\bibfnamefont {A.}~\bibnamefont
  {Ritz}},\ }\href {\doibase 10.1103/PhysRevD.84.075020} {\bibfield  {journal}
  {\bibinfo  {journal} {Phys. Rev. D}\ }\textbf {\bibinfo {volume} {84}},\
  \bibinfo {pages} {075020} (\bibinfo {year} {2011})},\ \Eprint
  {http://arxiv.org/abs/1107.4580} {arXiv:1107.4580 [hep-ph]} \BibitemShut
  {NoStop}%
\bibitem [{\citenamefont {Gninenko}(2012)}]{Gninenko:2012eq}%
  \BibitemOpen
  \bibfield  {author} {\bibinfo {author} {\bibfnamefont {S.~N.}\ \bibnamefont
  {Gninenko}},\ }\href {\doibase 10.1016/j.physletb.2012.06.002} {\bibfield
  {journal} {\bibinfo  {journal} {Phys. Lett. B}\ }\textbf {\bibinfo {volume}
  {713}},\ \bibinfo {pages} {244} (\bibinfo {year} {2012})},\ \Eprint
  {http://arxiv.org/abs/1204.3583} {arXiv:1204.3583 [hep-ph]} \BibitemShut
  {NoStop}%
\bibitem [{\citenamefont {Berlin}\ \emph {et~al.}(2018)\citenamefont {Berlin},
  \citenamefont {Gori}, \citenamefont {Schuster},\ and\ \citenamefont
  {Toro}}]{Berlin:2018pwi}%
  \BibitemOpen
  \bibfield  {author} {\bibinfo {author} {\bibfnamefont {A.}~\bibnamefont
  {Berlin}}, \bibinfo {author} {\bibfnamefont {S.}~\bibnamefont {Gori}},
  \bibinfo {author} {\bibfnamefont {P.}~\bibnamefont {Schuster}}, \ and\
  \bibinfo {author} {\bibfnamefont {N.}~\bibnamefont {Toro}},\ }\href {\doibase
  10.1103/PhysRevD.98.035011} {\bibfield  {journal} {\bibinfo  {journal} {Phys.
  Rev. D}\ }\textbf {\bibinfo {volume} {98}},\ \bibinfo {pages} {035011}
  (\bibinfo {year} {2018})},\ \Eprint {http://arxiv.org/abs/1804.00661}
  {arXiv:1804.00661 [hep-ph]} \BibitemShut {NoStop}%
\bibitem [{\citenamefont {Tsai}\ \emph {et~al.}(2021)\citenamefont {Tsai},
  \citenamefont {deNiverville},\ and\ \citenamefont {Liu}}]{Tsai:2019buq}%
  \BibitemOpen
  \bibfield  {author} {\bibinfo {author} {\bibfnamefont {Y.-D.}\ \bibnamefont
  {Tsai}}, \bibinfo {author} {\bibfnamefont {P.}~\bibnamefont {deNiverville}},
  \ and\ \bibinfo {author} {\bibfnamefont {M.~X.}\ \bibnamefont {Liu}},\ }\href
  {\doibase 10.1103/PhysRevLett.126.181801} {\bibfield  {journal} {\bibinfo
  {journal} {Phys. Rev. Lett.}\ }\textbf {\bibinfo {volume} {126}},\ \bibinfo
  {pages} {181801} (\bibinfo {year} {2021})},\ \Eprint
  {http://arxiv.org/abs/1908.07525} {arXiv:1908.07525 [hep-ph]} \BibitemShut
  {NoStop}%
\bibitem [{\citenamefont {Bruce}\ \emph {et~al.}(2020)\citenamefont {Bruce}
  \emph {et~al.}}]{Bruce:2018yzs}%
  \BibitemOpen
  \bibfield  {author} {\bibinfo {author} {\bibfnamefont {R.}~\bibnamefont
  {Bruce}} \emph {et~al.},\ }\href {\doibase 10.1088/1361-6471/ab7ff7}
  {\bibfield  {journal} {\bibinfo  {journal} {J. Phys. G}\ }\textbf {\bibinfo
  {volume} {47}},\ \bibinfo {pages} {060501} (\bibinfo {year} {2020})},\
  \Eprint {http://arxiv.org/abs/1812.07688} {arXiv:1812.07688 [hep-ph]}
  \BibitemShut {NoStop}%
\bibitem [{\citenamefont {Hayrapetyan}\ \emph {et~al.}(2023)\citenamefont
  {Hayrapetyan} \emph {et~al.}}]{CMS:2023thf}%
  \BibitemOpen
  \bibfield  {author} {\bibinfo {author} {\bibfnamefont {A.}~\bibnamefont
  {Hayrapetyan}} \emph {et~al.} (\bibinfo {collaboration} {CMS}),\ }\href
  {\doibase 10.1103/PhysRevLett.131.091903} {\bibfield  {journal} {\bibinfo
  {journal} {Phys. Rev. Lett.}\ }\textbf {\bibinfo {volume} {131}},\ \bibinfo
  {pages} {091903} (\bibinfo {year} {2023})},\ \Eprint
  {http://arxiv.org/abs/2305.04904} {arXiv:2305.04904 [hep-ex]} \BibitemShut
  {NoStop}%
\bibitem [{\citenamefont {Baur}\ and\ \citenamefont
  {Ferreira~Filho}(1990)}]{Baur:1990fx}%
  \BibitemOpen
  \bibfield  {author} {\bibinfo {author} {\bibfnamefont {G.}~\bibnamefont
  {Baur}}\ and\ \bibinfo {author} {\bibfnamefont {L.~G.}\ \bibnamefont
  {Ferreira~Filho}},\ }\href {\doibase 10.1016/0375-9474(90)90191-N} {\bibfield
   {journal} {\bibinfo  {journal} {Nucl. Phys. A}\ }\textbf {\bibinfo {volume}
  {518}},\ \bibinfo {pages} {786} (\bibinfo {year} {1990})}\BibitemShut
  {NoStop}%
\bibitem [{\citenamefont {Budnev}\ \emph {et~al.}(1975)\citenamefont {Budnev},
  \citenamefont {Ginzburg}, \citenamefont {Meledin},\ and\ \citenamefont
  {Serbo}}]{Budnev:1975poe}%
  \BibitemOpen
  \bibfield  {author} {\bibinfo {author} {\bibfnamefont {V.~M.}\ \bibnamefont
  {Budnev}}, \bibinfo {author} {\bibfnamefont {I.~F.}\ \bibnamefont
  {Ginzburg}}, \bibinfo {author} {\bibfnamefont {G.~V.}\ \bibnamefont
  {Meledin}}, \ and\ \bibinfo {author} {\bibfnamefont {V.~G.}\ \bibnamefont
  {Serbo}},\ }\href {\doibase 10.1016/0370-1573(75)90009-5} {\bibfield
  {journal} {\bibinfo  {journal} {Phys. Rept.}\ }\textbf {\bibinfo {volume}
  {15}},\ \bibinfo {pages} {181} (\bibinfo {year} {1975})}\BibitemShut
  {NoStop}%
\bibitem [{\citenamefont {Low}(1960)}]{Low:1960wv}%
  \BibitemOpen
  \bibfield  {author} {\bibinfo {author} {\bibfnamefont {F.~E.}\ \bibnamefont
  {Low}},\ }\href {\doibase 10.1103/PhysRev.120.582} {\bibfield  {journal}
  {\bibinfo  {journal} {Phys. Rev.}\ }\textbf {\bibinfo {volume} {120}},\
  \bibinfo {pages} {582} (\bibinfo {year} {1960})}\BibitemShut {NoStop}%
\bibitem [{\citenamefont {Klusek-Gawenda}\ and\ \citenamefont
  {Szczurek}(2010)}]{Klusek-Gawenda:2010vqb}%
  \BibitemOpen
  \bibfield  {author} {\bibinfo {author} {\bibfnamefont {M.}~\bibnamefont
  {Klusek-Gawenda}}\ and\ \bibinfo {author} {\bibfnamefont {A.}~\bibnamefont
  {Szczurek}},\ }\href {\doibase 10.1103/PhysRevC.82.014904} {\bibfield
  {journal} {\bibinfo  {journal} {Phys. Rev. C}\ }\textbf {\bibinfo {volume}
  {82}},\ \bibinfo {pages} {014904} (\bibinfo {year} {2010})},\ \Eprint
  {http://arxiv.org/abs/1004.5521} {arXiv:1004.5521 [nucl-th]} \BibitemShut
  {NoStop}%
\bibitem [{\citenamefont {Hencken}\ \emph {et~al.}(1995)\citenamefont
  {Hencken}, \citenamefont {Trautmann},\ and\ \citenamefont
  {Baur}}]{Hencken:1995me}%
  \BibitemOpen
  \bibfield  {author} {\bibinfo {author} {\bibfnamefont {K.}~\bibnamefont
  {Hencken}}, \bibinfo {author} {\bibfnamefont {D.}~\bibnamefont {Trautmann}},
  \ and\ \bibinfo {author} {\bibfnamefont {G.}~\bibnamefont {Baur}},\ }\href
  {\doibase 10.1007/BF01620724} {\bibfield  {journal} {\bibinfo  {journal} {Z.
  Phys. C}\ }\textbf {\bibinfo {volume} {68}},\ \bibinfo {pages} {473}
  (\bibinfo {year} {1995})},\ \Eprint {http://arxiv.org/abs/nucl-th/9503004}
  {arXiv:nucl-th/9503004} \BibitemShut {NoStop}%
\bibitem [{\citenamefont {Goncalves}(2013)}]{Goncalves:2012cy}%
  \BibitemOpen
  \bibfield  {author} {\bibinfo {author} {\bibfnamefont {V.~P.}\ \bibnamefont
  {Goncalves}},\ }\href {\doibase 10.1016/j.nuclphysa.2013.02.016} {\bibfield
  {journal} {\bibinfo  {journal} {Nucl. Phys. A}\ }\textbf {\bibinfo {volume}
  {902}},\ \bibinfo {pages} {32} (\bibinfo {year} {2013})},\ \Eprint
  {http://arxiv.org/abs/1211.1207} {arXiv:1211.1207 [hep-ph]} \BibitemShut
  {NoStop}%
\bibitem [{\citenamefont {Goncalves}\ and\ \citenamefont
  {Sauter}(2015)}]{Goncalves:2015hra}%
  \BibitemOpen
  \bibfield  {author} {\bibinfo {author} {\bibfnamefont {V.~P.}\ \bibnamefont
  {Goncalves}}\ and\ \bibinfo {author} {\bibfnamefont {W.~K.}\ \bibnamefont
  {Sauter}},\ }\href {\doibase 10.1103/PhysRevD.91.094014} {\bibfield
  {journal} {\bibinfo  {journal} {Phys. Rev. D}\ }\textbf {\bibinfo {volume}
  {91}},\ \bibinfo {pages} {094014} (\bibinfo {year} {2015})},\ \Eprint
  {http://arxiv.org/abs/1503.05112} {arXiv:1503.05112 [hep-ph]} \BibitemShut
  {NoStop}%
\bibitem [{\citenamefont {Gon\c{c}alves}\ and\ \citenamefont
  {Moreira}(2018)}]{Goncalves:2018yxc}%
  \BibitemOpen
  \bibfield  {author} {\bibinfo {author} {\bibfnamefont {V.~P.}\ \bibnamefont
  {Gon\c{c}alves}}\ and\ \bibinfo {author} {\bibfnamefont {B.~D.}\ \bibnamefont
  {Moreira}},\ }\href {\doibase 10.1103/PhysRevD.97.094009} {\bibfield
  {journal} {\bibinfo  {journal} {Phys. Rev. D}\ }\textbf {\bibinfo {volume}
  {97}},\ \bibinfo {pages} {094009} (\bibinfo {year} {2018})},\ \Eprint
  {http://arxiv.org/abs/1801.10501} {arXiv:1801.10501 [hep-ph]} \BibitemShut
  {NoStop}%
\bibitem [{\citenamefont {Krovi}\ \emph {et~al.}(2020)\citenamefont {Krovi},
  \citenamefont {Low},\ and\ \citenamefont {Zhang}}]{Krovi:2019hdl}%
  \BibitemOpen
  \bibfield  {author} {\bibinfo {author} {\bibfnamefont {A.}~\bibnamefont
  {Krovi}}, \bibinfo {author} {\bibfnamefont {I.}~\bibnamefont {Low}}, \ and\
  \bibinfo {author} {\bibfnamefont {Y.}~\bibnamefont {Zhang}},\ }\href
  {\doibase 10.1103/PhysRevD.102.055003} {\bibfield  {journal} {\bibinfo
  {journal} {Phys. Rev. D}\ }\textbf {\bibinfo {volume} {102}},\ \bibinfo
  {pages} {055003} (\bibinfo {year} {2020})},\ \Eprint
  {http://arxiv.org/abs/1909.07987} {arXiv:1909.07987 [hep-ph]} \BibitemShut
  {NoStop}%
\bibitem [{\citenamefont {Buschmann}\ \emph {et~al.}(2015)\citenamefont
  {Buschmann}, \citenamefont {Kopp}, \citenamefont {Liu},\ and\ \citenamefont
  {Machado}}]{Buschmann:2015awa}%
  \BibitemOpen
  \bibfield  {author} {\bibinfo {author} {\bibfnamefont {M.}~\bibnamefont
  {Buschmann}}, \bibinfo {author} {\bibfnamefont {J.}~\bibnamefont {Kopp}},
  \bibinfo {author} {\bibfnamefont {J.}~\bibnamefont {Liu}}, \ and\ \bibinfo
  {author} {\bibfnamefont {P.~A.~N.}\ \bibnamefont {Machado}},\ }\href
  {\doibase 10.1007/JHEP07(2015)045} {\bibfield  {journal} {\bibinfo  {journal}
  {JHEP}\ }\textbf {\bibinfo {volume} {07}},\ \bibinfo {pages} {045} (\bibinfo
  {year} {2015})},\ \Eprint {http://arxiv.org/abs/1505.07459} {arXiv:1505.07459
  [hep-ph]} \BibitemShut {NoStop}%
\bibitem [{\citenamefont {Raggi}\ and\ \citenamefont
  {Kozhuharov}(2015)}]{Raggi:2015yfk}%
  \BibitemOpen
  \bibfield  {author} {\bibinfo {author} {\bibfnamefont {M.}~\bibnamefont
  {Raggi}}\ and\ \bibinfo {author} {\bibfnamefont {V.}~\bibnamefont
  {Kozhuharov}},\ }\href {\doibase 10.1393/ncr/i2015-10117-9} {\bibfield
  {journal} {\bibinfo  {journal} {Riv. Nuovo Cim.}\ }\textbf {\bibinfo {volume}
  {38}},\ \bibinfo {pages} {449} (\bibinfo {year} {2015})}\BibitemShut
  {NoStop}%
\bibitem [{\citenamefont {Fabbrichesi}\ \emph {et~al.}(2020)\citenamefont
  {Fabbrichesi}, \citenamefont {Gabrielli},\ and\ \citenamefont
  {Lanfranchi}}]{Fabbrichesi:2020wbt}%
  \BibitemOpen
  \bibfield  {author} {\bibinfo {author} {\bibfnamefont {M.}~\bibnamefont
  {Fabbrichesi}}, \bibinfo {author} {\bibfnamefont {E.}~\bibnamefont
  {Gabrielli}}, \ and\ \bibinfo {author} {\bibfnamefont {G.}~\bibnamefont
  {Lanfranchi}},\ }\href {\doibase 10.1007/978-3-030-62519-1} {\  (\bibinfo
  {year} {2020}),\ 10.1007/978-3-030-62519-1},\ \Eprint
  {http://arxiv.org/abs/2005.01515} {arXiv:2005.01515 [hep-ph]} \BibitemShut
  {NoStop}%
\bibitem [{\citenamefont {Okun}(1982)}]{okun1982limits}%
  \BibitemOpen
  \bibfield  {author} {\bibinfo {author} {\bibfnamefont {L.~B.}\ \bibnamefont
  {Okun}},\ }\href@noop {} {\bibfield  {journal} {\bibinfo  {journal} {Sov.
  Phys.-JETP}\ }\textbf {\bibinfo {volume} {56}},\ \bibinfo {pages} {502}
  (\bibinfo {year} {1982})}\BibitemShut {NoStop}%
\bibitem [{\citenamefont {Holdom}(1986)}]{holdom1986two}%
  \BibitemOpen
  \bibfield  {author} {\bibinfo {author} {\bibfnamefont {B.}~\bibnamefont
  {Holdom}},\ }\href@noop {} {\bibfield  {journal} {\bibinfo  {journal} {Phys.
  Lett. B;(Netherlands)}\ }\textbf {\bibinfo {volume} {166}} (\bibinfo {year}
  {1986})}\BibitemShut {NoStop}%
\bibitem [{\citenamefont {Foot}\ and\ \citenamefont {He}(1991)}]{Foot:1991kb}%
  \BibitemOpen
  \bibfield  {author} {\bibinfo {author} {\bibfnamefont {R.}~\bibnamefont
  {Foot}}\ and\ \bibinfo {author} {\bibfnamefont {X.-G.}\ \bibnamefont {He}},\
  }\href {\doibase 10.1016/0370-2693(91)90901-2} {\bibfield  {journal}
  {\bibinfo  {journal} {Phys. Lett. B}\ }\textbf {\bibinfo {volume} {267}},\
  \bibinfo {pages} {509} (\bibinfo {year} {1991})}\BibitemShut {NoStop}%
\bibitem [{\citenamefont {Jaeckel}\ and\ \citenamefont
  {Ringwald}(2010)}]{Jaeckel:2010ni}%
  \BibitemOpen
  \bibfield  {author} {\bibinfo {author} {\bibfnamefont {J.}~\bibnamefont
  {Jaeckel}}\ and\ \bibinfo {author} {\bibfnamefont {A.}~\bibnamefont
  {Ringwald}},\ }\href {\doibase 10.1146/annurev.nucl.012809.104433} {\bibfield
   {journal} {\bibinfo  {journal} {Ann. Rev. Nucl. Part. Sci.}\ }\textbf
  {\bibinfo {volume} {60}},\ \bibinfo {pages} {405} (\bibinfo {year} {2010})},\
  \Eprint {http://arxiv.org/abs/1002.0329} {arXiv:1002.0329 [hep-ph]}
  \BibitemShut {NoStop}%
\bibitem [{\citenamefont {Bergsma}\ \emph {et~al.}(1986)\citenamefont {Bergsma}
  \emph {et~al.}}]{CHARM:1985nku}%
  \BibitemOpen
  \bibfield  {author} {\bibinfo {author} {\bibfnamefont {F.}~\bibnamefont
  {Bergsma}} \emph {et~al.} (\bibinfo {collaboration} {CHARM}),\ }\href
  {\doibase 10.1016/0370-2693(86)91601-1} {\bibfield  {journal} {\bibinfo
  {journal} {Phys. Lett. B}\ }\textbf {\bibinfo {volume} {166}},\ \bibinfo
  {pages} {473} (\bibinfo {year} {1986})}\BibitemShut {NoStop}%
\bibitem [{\citenamefont {Pospelov}\ \emph {et~al.}(2008)\citenamefont
  {Pospelov}, \citenamefont {Ritz},\ and\ \citenamefont
  {Voloshin}}]{Pospelov:2007mp}%
  \BibitemOpen
  \bibfield  {author} {\bibinfo {author} {\bibfnamefont {M.}~\bibnamefont
  {Pospelov}}, \bibinfo {author} {\bibfnamefont {A.}~\bibnamefont {Ritz}}, \
  and\ \bibinfo {author} {\bibfnamefont {M.~B.}\ \bibnamefont {Voloshin}},\
  }\href {\doibase 10.1016/j.physletb.2008.02.052} {\bibfield  {journal}
  {\bibinfo  {journal} {Phys. Lett. B}\ }\textbf {\bibinfo {volume} {662}},\
  \bibinfo {pages} {53} (\bibinfo {year} {2008})},\ \Eprint
  {http://arxiv.org/abs/0711.4866} {arXiv:0711.4866 [hep-ph]} \BibitemShut
  {NoStop}%
\bibitem [{\citenamefont {Ilten}\ \emph {et~al.}(2015)\citenamefont {Ilten},
  \citenamefont {Thaler}, \citenamefont {Williams},\ and\ \citenamefont
  {Xue}}]{Ilten:2015hya}%
  \BibitemOpen
  \bibfield  {author} {\bibinfo {author} {\bibfnamefont {P.}~\bibnamefont
  {Ilten}}, \bibinfo {author} {\bibfnamefont {J.}~\bibnamefont {Thaler}},
  \bibinfo {author} {\bibfnamefont {M.}~\bibnamefont {Williams}}, \ and\
  \bibinfo {author} {\bibfnamefont {W.}~\bibnamefont {Xue}},\ }\href {\doibase
  10.1103/PhysRevD.92.115017} {\bibfield  {journal} {\bibinfo  {journal} {Phys.
  Rev. D}\ }\textbf {\bibinfo {volume} {92}},\ \bibinfo {pages} {115017}
  (\bibinfo {year} {2015})},\ \Eprint {http://arxiv.org/abs/1509.06765}
  {arXiv:1509.06765 [hep-ph]} \BibitemShut {NoStop}%
\bibitem [{\citenamefont {Workman}\ and\ \citenamefont
  {Others}(2022)}]{Workman:2022ynf}%
  \BibitemOpen
  \bibfield  {author} {\bibinfo {author} {\bibfnamefont {R.~L.}\ \bibnamefont
  {Workman}}\ and\ \bibinfo {author} {\bibnamefont {Others}} (\bibinfo
  {collaboration} {Particle Data Group}),\ }\href {\doibase
  10.1093/ptep/ptac097} {\bibfield  {journal} {\bibinfo  {journal} {PTEP}\
  }\textbf {\bibinfo {volume} {2022}},\ \bibinfo {pages} {083C01} (\bibinfo
  {year} {2022})}\BibitemShut {NoStop}%
\bibitem [{\citenamefont {Fayet}(2007)}]{Fayet:2007ua}%
  \BibitemOpen
  \bibfield  {author} {\bibinfo {author} {\bibfnamefont {P.}~\bibnamefont
  {Fayet}},\ }\href {\doibase 10.1103/PhysRevD.75.115017} {\bibfield  {journal}
  {\bibinfo  {journal} {Phys. Rev. D}\ }\textbf {\bibinfo {volume} {75}},\
  \bibinfo {pages} {115017} (\bibinfo {year} {2007})},\ \Eprint
  {http://arxiv.org/abs/hep-ph/0702176} {arXiv:hep-ph/0702176} \BibitemShut
  {NoStop}%
\bibitem [{\citenamefont {Pospelov}(2009)}]{Pospelov:2008zw}%
  \BibitemOpen
  \bibfield  {author} {\bibinfo {author} {\bibfnamefont {M.}~\bibnamefont
  {Pospelov}},\ }\href {\doibase 10.1103/PhysRevD.80.095002} {\bibfield
  {journal} {\bibinfo  {journal} {Phys. Rev. D}\ }\textbf {\bibinfo {volume}
  {80}},\ \bibinfo {pages} {095002} (\bibinfo {year} {2009})},\ \Eprint
  {http://arxiv.org/abs/0811.1030} {arXiv:0811.1030 [hep-ph]} \BibitemShut
  {NoStop}%
\bibitem [{\citenamefont {Bennett}\ \emph {et~al.}(2006)\citenamefont {Bennett}
  \emph {et~al.}}]{Muong-2:2006rrc}%
  \BibitemOpen
  \bibfield  {author} {\bibinfo {author} {\bibfnamefont {G.~W.}\ \bibnamefont
  {Bennett}} \emph {et~al.} (\bibinfo {collaboration} {Muon g-2}),\ }\href
  {\doibase 10.1103/PhysRevD.73.072003} {\bibfield  {journal} {\bibinfo
  {journal} {Phys. Rev. D}\ }\textbf {\bibinfo {volume} {73}},\ \bibinfo
  {pages} {072003} (\bibinfo {year} {2006})},\ \Eprint
  {http://arxiv.org/abs/hep-ex/0602035} {arXiv:hep-ex/0602035} \BibitemShut
  {NoStop}%
\bibitem [{\citenamefont {Aoyama}\ \emph {et~al.}(2020)\citenamefont {Aoyama}
  \emph {et~al.}}]{Aoyama:2020ynm}%
  \BibitemOpen
  \bibfield  {author} {\bibinfo {author} {\bibfnamefont {T.}~\bibnamefont
  {Aoyama}} \emph {et~al.},\ }\href {\doibase 10.1016/j.physrep.2020.07.006}
  {\bibfield  {journal} {\bibinfo  {journal} {Phys. Rept.}\ }\textbf {\bibinfo
  {volume} {887}},\ \bibinfo {pages} {1} (\bibinfo {year} {2020})},\ \Eprint
  {http://arxiv.org/abs/2006.04822} {arXiv:2006.04822 [hep-ph]} \BibitemShut
  {NoStop}%
\bibitem [{\citenamefont {Abi}\ \emph {et~al.}(2021)\citenamefont {Abi} \emph
  {et~al.}}]{Muong-2:2021ojo}%
  \BibitemOpen
  \bibfield  {author} {\bibinfo {author} {\bibfnamefont {B.}~\bibnamefont
  {Abi}} \emph {et~al.} (\bibinfo {collaboration} {Muon g-2}),\ }\href
  {\doibase 10.1103/PhysRevLett.126.141801} {\bibfield  {journal} {\bibinfo
  {journal} {Phys. Rev. Lett.}\ }\textbf {\bibinfo {volume} {126}},\ \bibinfo
  {pages} {141801} (\bibinfo {year} {2021})},\ \Eprint
  {http://arxiv.org/abs/2104.03281} {arXiv:2104.03281 [hep-ex]} \BibitemShut
  {NoStop}%
\bibitem [{\citenamefont {Albahri}\ \emph {et~al.}(2021)\citenamefont {Albahri}
  \emph {et~al.}}]{Muong-2:2021vma}%
  \BibitemOpen
  \bibfield  {author} {\bibinfo {author} {\bibfnamefont {T.}~\bibnamefont
  {Albahri}} \emph {et~al.} (\bibinfo {collaboration} {Muon g-2}),\ }\href
  {\doibase 10.1103/PhysRevD.103.072002} {\bibfield  {journal} {\bibinfo
  {journal} {Phys. Rev. D}\ }\textbf {\bibinfo {volume} {103}},\ \bibinfo
  {pages} {072002} (\bibinfo {year} {2021})},\ \Eprint
  {http://arxiv.org/abs/2104.03247} {arXiv:2104.03247 [hep-ex]} \BibitemShut
  {NoStop}%
\bibitem [{\citenamefont {Abada}\ \emph
  {et~al.}(2019{\natexlab{a}})\citenamefont {Abada} \emph
  {et~al.}}]{FCC:2018vvp}%
  \BibitemOpen
  \bibfield  {author} {\bibinfo {author} {\bibfnamefont {A.}~\bibnamefont
  {Abada}} \emph {et~al.} (\bibinfo {collaboration} {FCC}),\ }\href {\doibase
  10.1140/epjst/e2019-900087-0} {\bibfield  {journal} {\bibinfo  {journal}
  {Eur. Phys. J. ST}\ }\textbf {\bibinfo {volume} {228}},\ \bibinfo {pages}
  {755} (\bibinfo {year} {2019}{\natexlab{a}})}\BibitemShut {NoStop}%
\bibitem [{\citenamefont {Abada}\ \emph
  {et~al.}(2019{\natexlab{b}})\citenamefont {Abada} \emph
  {et~al.}}]{FCC:2018bvk}%
  \BibitemOpen
  \bibfield  {author} {\bibinfo {author} {\bibfnamefont {A.}~\bibnamefont
  {Abada}} \emph {et~al.} (\bibinfo {collaboration} {FCC}),\ }\href {\doibase
  10.1140/epjst/e2019-900088-6} {\bibfield  {journal} {\bibinfo  {journal}
  {Eur. Phys. J. ST}\ }\textbf {\bibinfo {volume} {228}},\ \bibinfo {pages}
  {1109} (\bibinfo {year} {2019}{\natexlab{b}})}\BibitemShut {NoStop}%
\bibitem [{\citenamefont {Abada}\ \emph
  {et~al.}(2019{\natexlab{c}})\citenamefont {Abada} \emph
  {et~al.}}]{FCC:2018byv}%
  \BibitemOpen
  \bibfield  {author} {\bibinfo {author} {\bibfnamefont {A.}~\bibnamefont
  {Abada}} \emph {et~al.} (\bibinfo {collaboration} {FCC}),\ }\href {\doibase
  10.1140/epjc/s10052-019-6904-3} {\bibfield  {journal} {\bibinfo  {journal}
  {Eur. Phys. J. C}\ }\textbf {\bibinfo {volume} {79}},\ \bibinfo {pages} {474}
  (\bibinfo {year} {2019}{\natexlab{c}})}\BibitemShut {NoStop}%
\bibitem [{\citenamefont {Soyez}(2019)}]{Soyez:2018opl}%
  \BibitemOpen
  \bibfield  {author} {\bibinfo {author} {\bibfnamefont {G.}~\bibnamefont
  {Soyez}},\ }\href {\doibase 10.1016/j.physrep.2019.01.007} {\bibfield
  {journal} {\bibinfo  {journal} {Phys. Rept.}\ }\textbf {\bibinfo {volume}
  {803}},\ \bibinfo {pages} {1} (\bibinfo {year} {2019})},\ \Eprint
  {http://arxiv.org/abs/1801.09721} {arXiv:1801.09721 [hep-ph]} \BibitemShut
  {NoStop}%
\bibitem [{\citenamefont {Choi}\ \emph {et~al.}(2019)\citenamefont {Choi},
  \citenamefont {Ryu},\ and\ \citenamefont {Ji}}]{Choi:2019wqx}%
  \BibitemOpen
  \bibfield  {author} {\bibinfo {author} {\bibfnamefont {H.-M.}\ \bibnamefont
  {Choi}}, \bibinfo {author} {\bibfnamefont {H.-Y.}\ \bibnamefont {Ryu}}, \
  and\ \bibinfo {author} {\bibfnamefont {C.-R.}\ \bibnamefont {Ji}},\ }\href
  {\doibase 10.1103/PhysRevD.99.076012} {\bibfield  {journal} {\bibinfo
  {journal} {Phys. Rev. D}\ }\textbf {\bibinfo {volume} {99}},\ \bibinfo
  {pages} {076012} (\bibinfo {year} {2019})},\ \Eprint
  {http://arxiv.org/abs/1903.01448} {arXiv:1903.01448 [hep-ph]} \BibitemShut
  {NoStop}%
\bibitem [{\citenamefont {Elam}\ \emph {et~al.}(2022)\citenamefont {Elam} \emph
  {et~al.}}]{REDTOP:2022slw}%
  \BibitemOpen
  \bibfield  {author} {\bibinfo {author} {\bibfnamefont {J.}~\bibnamefont
  {Elam}} \emph {et~al.} (\bibinfo {collaboration} {REDTOP}),\ }\href@noop {}
  {\  (\bibinfo {year} {2022})},\ \Eprint {http://arxiv.org/abs/2203.07651}
  {arXiv:2203.07651 [hep-ex]} \BibitemShut {NoStop}%
\bibitem [{\citenamefont {Primakoff}(1951)}]{Primakoff:1951iae}%
  \BibitemOpen
  \bibfield  {author} {\bibinfo {author} {\bibfnamefont {H.}~\bibnamefont
  {Primakoff}},\ }\href {\doibase 10.1103/PhysRev.81.899} {\bibfield  {journal}
  {\bibinfo  {journal} {Phys. Rev.}\ }\textbf {\bibinfo {volume} {81}},\
  \bibinfo {pages} {899} (\bibinfo {year} {1951})}\BibitemShut {NoStop}%
\bibitem [{\citenamefont {Accardi}\ \emph {et~al.}(2016)\citenamefont {Accardi}
  \emph {et~al.}}]{Accardi:2012qut}%
  \BibitemOpen
  \bibfield  {author} {\bibinfo {author} {\bibfnamefont {A.}~\bibnamefont
  {Accardi}} \emph {et~al.},\ }\href {\doibase 10.1140/epja/i2016-16268-9}
  {\bibfield  {journal} {\bibinfo  {journal} {Eur. Phys. J. A}\ }\textbf
  {\bibinfo {volume} {52}},\ \bibinfo {pages} {268} (\bibinfo {year} {2016})},\
  \Eprint {http://arxiv.org/abs/1212.1701} {arXiv:1212.1701 [nucl-ex]}
  \BibitemShut {NoStop}%
\bibitem [{\citenamefont {Abdul~Khalek}\ \emph {et~al.}(2022)\citenamefont
  {Abdul~Khalek} \emph {et~al.}}]{AbdulKhalek:2021gbh}%
  \BibitemOpen
  \bibfield  {author} {\bibinfo {author} {\bibfnamefont {R.}~\bibnamefont
  {Abdul~Khalek}} \emph {et~al.},\ }\href {\doibase
  10.1016/j.nuclphysa.2022.122447} {\bibfield  {journal} {\bibinfo  {journal}
  {Nucl. Phys. A}\ }\textbf {\bibinfo {volume} {1026}},\ \bibinfo {pages}
  {122447} (\bibinfo {year} {2022})},\ \Eprint
  {http://arxiv.org/abs/2103.05419} {arXiv:2103.05419 [physics.ins-det]}
  \BibitemShut {NoStop}%
\bibitem [{\citenamefont {Chen}(2018)}]{Chen:2018wyz}%
  \BibitemOpen
  \bibfield  {author} {\bibinfo {author} {\bibfnamefont {X.}~\bibnamefont
  {Chen}},\ }\href {\doibase 10.22323/1.316.0170} {\bibfield  {journal}
  {\bibinfo  {journal} {PoS}\ }\textbf {\bibinfo {volume} {DIS2018}},\ \bibinfo
  {pages} {170} (\bibinfo {year} {2018})},\ \Eprint
  {http://arxiv.org/abs/1809.00448} {arXiv:1809.00448 [nucl-ex]} \BibitemShut
  {NoStop}%
\bibitem [{\citenamefont {Chen}\ \emph {et~al.}(2020)\citenamefont {Chen},
  \citenamefont {Guo}, \citenamefont {Roberts},\ and\ \citenamefont
  {Wang}}]{Chen:2020ijn}%
  \BibitemOpen
  \bibfield  {author} {\bibinfo {author} {\bibfnamefont {X.}~\bibnamefont
  {Chen}}, \bibinfo {author} {\bibfnamefont {F.-K.}\ \bibnamefont {Guo}},
  \bibinfo {author} {\bibfnamefont {C.~D.}\ \bibnamefont {Roberts}}, \ and\
  \bibinfo {author} {\bibfnamefont {R.}~\bibnamefont {Wang}},\ }\href {\doibase
  10.1007/s00601-020-01574-0} {\bibfield  {journal} {\bibinfo  {journal} {Few
  Body Syst.}\ }\textbf {\bibinfo {volume} {61}},\ \bibinfo {pages} {43}
  (\bibinfo {year} {2020})},\ \Eprint {http://arxiv.org/abs/2008.00102}
  {arXiv:2008.00102 [hep-ph]} \BibitemShut {NoStop}%
\bibitem [{\citenamefont {Anderle}\ \emph {et~al.}(2021)\citenamefont {Anderle}
  \emph {et~al.}}]{Anderle:2021wcy}%
  \BibitemOpen
  \bibfield  {author} {\bibinfo {author} {\bibfnamefont {D.~P.}\ \bibnamefont
  {Anderle}} \emph {et~al.},\ }\href {\doibase 10.1007/s11467-021-1062-0}
  {\bibfield  {journal} {\bibinfo  {journal} {Front. Phys. (Beijing)}\ }\textbf
  {\bibinfo {volume} {16}},\ \bibinfo {pages} {64701} (\bibinfo {year}
  {2021})},\ \Eprint {http://arxiv.org/abs/2102.09222} {arXiv:2102.09222
  [nucl-ex]} \BibitemShut {NoStop}%
\end{thebibliography}%
\newpage

\end{document}